\pdfoutput=1
\documentclass[11pt,letterpaper]{article}

\usepackage[margin=1in]{geometry}
\usepackage[T1]{fontenc}
\usepackage{lmodern}
\usepackage[english]{babel}
\usepackage{microtype}
\usepackage{comment}
\usepackage{setspace}
\usepackage{mathtools}  
\usepackage{amssymb}  
\usepackage{amsthm}
\usepackage{bm}
\usepackage{mathrsfs}
\numberwithin{equation}{section}

\usepackage{placeins} 
\usepackage{graphicx}
\usepackage{booktabs}
\usepackage{multirow}
\usepackage{makecell}
\usepackage{adjustbox}
\usepackage[flushleft]{threeparttable}
\usepackage{rotating}
\usepackage{pdflscape}
\usepackage{afterpage}
\usepackage{algorithm}
\usepackage{algpseudocode}  
\usepackage[font=small,labelfont=bf]{caption}
\usepackage{subcaption}
\usepackage[shortlabels]{enumitem}
\usepackage[title]{appendix}
\usepackage{xcolor}
\usepackage{authblk}

\usepackage[authoryear,round]{natbib}
\theoremstyle{plain}
\newtheorem{theorem}{Theorem}[section]
\newtheorem{lemma}[theorem]{Lemma}
\newtheorem{proposition}[theorem]{Proposition}
\newtheorem{corollary}[theorem]{Corollary}
\theoremstyle{definition}

\theoremstyle{remark}

\theoremstyle{plain}
\newcommand{\restatename}{}
\newtheorem*{restate}{\restatename}

\newcommand{\sym}[1]{\ifmmode^{#1}\else\(^{#1}\)\fi}

\usepackage[colorlinks=true,linkcolor=black,citecolor=black,urlcolor=blue]{hyperref}
\graphicspath{{figures/}}

\begin{document}

\thispagestyle{empty}
\vspace*{0.6cm}

\begin{center}
    {\fontsize{22}{26}\selectfont
    \bfseries
    Differential Privacy for Generalized Entropy Indices of Income Inequality
    \par}
\end{center}

\begin{center}
    {\large Jiahe Li\par}
    {\normalsize Department of Statistical Science, Duke University\par}
    {\normalsize and\par}
    {\large Jerome P. Reiter\par}
    {\normalsize Department of Statistical Science, Duke University\par}
    {\large \today\par}
\end{center}
\vspace{1.0cm}
\begin{abstract}
\noindent
The mean log deviation and the Theil index are two widely used 
measures of income inequality.
The organization releasing these measures may be obligated to protect the confidentiality of the individual incomes used in the computations.  In particular, the organization may wish to release versions of these measures with formal guarantees of privacy.  The organization could do so by satisfying differential privacy; however, typical applications of differentially private algorithms are likely to add so much noise to the measures that they become unusable. We present differentially private algorithms for these indices with reasonable error properties.  The algorithms rely on computing or approximating bounds on the smooth sensitivity, that is, the maximum absolute amount the inequality measures change when replacing any subset of the $n$ incomes in the reported data with  arbitrarily chosen feasible values. 
For the mean log deviation, the algorithm runs in 
$O(n^2)$ time. For the Theil index, the algorithm runs 
in $O(n\log n)$ time. 
We illustrate the performance of each differentially private algorithm using individual-level and household-level income data from the Current Population Survey. We also examine repeated sampling properties of the algorithms using artificially generated data. The empirical results indicate when these differentially private algorithms can be expected to provide accurate estimates  and when they are likely to introduce too much noise.
\end{abstract}

\vspace{0.2cm}

\noindent
\textit{Keywords}:  Confidentiality; Mean Log Deviation; Smooth Sensitivity; Theil Index.

\newpage

\section{Introduction}
\label{sec:introduction}

Income inequality is commonly characterized by 
the class of generalized entropy  measures 
\citep{bourguignon1979decomposable,cowell1980structure,shorrocks1980class,shorrocks1984inequality}. 
This class is indexed by a parameter that determines the relative emphasis placed on different parts of the income distribution. Smaller values of the parameter emphasize differences in the lower tail of the distribution, and larger values emphasize differences in the upper tail. The two most widely used generalized entropy measures of income inequality include the limiting cases when the indexing parameter equals zero, known as the mean log deviation (MLD), and when the indexing parameter equals one, known as the Theil index.  Both inequality measures were introduced by \citet{theil1967economics}. Statistical agencies often release one or both measures when 
summarizing income inequality.
For example, the U.S.\ Census Bureau reports the MLD and the Theil index in its annual income report. In 2024,  the reported MLD and the Theil indices for household income were 0.606 and 0.437, respectively \citep{kollar2025income}.

The microdata underlying these inequality measures contain sensitive information, i.e., incomes, about individuals and households. Typically, the organizations charged with collecting the microdata and disseminating the inequality measures  are 
obligated to protect the confidentiality of this information. For example, data collected by the Census Bureau 
are confidential under Title 13 and, for data provided to the Census Bureau by the Internal Revenue Service, Title 26 of the U.S.\ code. Thus, organizations may wish to limit the amount of information that external users can learn about individual incomes from their released data products.


One way to limit the amount of information leakage about individuals' data values is to release statistics that satisfy differential privacy (DP) 
\citep{dwork2006calibrating,dwork2006differential,dwork2014algorithmic}. DP offers mathematically provable guarantees of privacy protection. Because of its appealing features, several organizations in the private and public sector are designing data release strategies that satisfy DP \citep{topdown, BI2023444}.   
The Census Bureau releases differentially private earnings percentiles of college graduates in its Post-Secondary Employment Outcomes data \citep{foote2019releasing}. \citet{barrientos2024feasibility} evaluate DP methods on tax data from the Internal Revenue Service and find them feasible for univariate summary statistics. In the private sector, LinkedIn uses differential privacy to provide advertisers with analytics on member engagement while offering user-level privacy guarantees \citep{rogers2021linkedin}.
Meta and Opportunity Insights add noise based on methods from the DP literature to release measures of economic connectedness between individuals with low and high socioeconomic status \citep{chetty2022social}.

To satisfy DP, a typical strategy is to add random noise to the statistic.  The variance of the noise distribution is calibrated to what is known as the global sensitivity, defined as the largest possible change in the statistic over all hypothetical databases in the domain caused by altering a single record. Because global sensitivity is determined by worst-case scenarios, it can result in a large noise variance to guarantee privacy protection, which in turn can make the released noisy statistic inaccurate. One way to avoid this variance inflation is to design DP algorithms to satisfy what is known as smooth sensitivity  \citep{nissim2007smooth}, which allows the noise scale to adapt to the observed data at hand while ensuring DP.

In this article, we use smooth sensitivity to develop algorithms for releasing DP versions of the MLD and Theil index. We first derive expressions for the global sensitivity of each index.  We also derive  expressions for the local sensitivity of each index, that is, the maximum each index can change when altering a single income in the data at hand. 
For the MLD, we exactly characterize the maximum local sensitivity over every possible dataset differing in $k$ elements, where $k$ ranges from one to the sample size $n$. These characterizations enable computation of the smooth sensitivity in $O(n^2)$ time. For the Theil index, we construct a computable smooth upper bound on the local sensitivity, which enables computation of the smooth sensitivity in $O(n\log n)$ time.  We illustrate the performance of the DP measures using individual and household income data from the Current Population Survey.  We also examine repeated sampling performances of the DP algorithms using artificially generated data.  These simulations suggest what types of  sample sizes and privacy requirements 
reliably admit accurate  DP inequality measures, as well as what types do  not. 

 To our knowledge, DP algorithms for the MLD and Theil index have not been developed.  The most closely related work to ours is that of \citet{lan2026differentially}. They use smooth sensitivity to design a DP mechanism for the Gini index of inequality, achieving 
improved accuracy over mechanisms based on global sensitivity.  Our algorithms differ from those of \citet{lan2026differentially} in that they are specifically tailored to releasing the MLD and Theil index.

The remainder of this article is organized as follows. Section~\ref{sec:background} reviews the MLD and Theil index, differential privacy, and smooth sensitivity. Section~\ref{sec:dp-mld} and Section~\ref{sec:dp-theil} develop the sensitivity results and DP release mechanisms for the MLD and the Theil index, respectively. Section~\ref{sec:experiments} evaluates the algorithms using income data from the Current Population Survey (CPS) and using artificially generated data. Section~\ref{sec:discussion} concludes with suggestions for future extensions.  Code for all methods and simulations are available at \href{https://github.com/JiaheLee/DP_Generalized_Entropy}{\url{https://github.com/JiaheLee/DP_Generalized_Entropy}}.

\section{Background}
\label{sec:background}

We first review generalized entropy indices  followed by reviews of differential privacy and smooth sensitivity.

\subsection{Generalized Entropy Indices}
\label{subsec:ge-background}

For $i=1, \dots, n$ individuals, let $x_i$ be each individual's
income, and let  $X=(x_1,\ldots,x_n)$. 
Let $\bar x= \sum_{i=1}^n x_i/n$ denote the sample mean.
For $a\in\mathbb R\setminus\{0,1\}$, the generalized entropy  index is
\begin{equation}
\label{eq:ge-definition}
    \operatorname{GE}_a(X)
    =\frac{1}{a(a-1)}
      \left\{
        \frac1n\sum_{i=1}^n
        \left(\frac{x_i}{\bar x}\right)^a-1
      \right\}.
\end{equation}
The indices at $a=0$ and $a=1$ are defined by continuity.
The mean log deviation  is
\begin{equation}
\label{eq:mld-definition}
    \operatorname{GE}_0(X) =  M(X) 
    =\frac1n\sum_{i=1}^n
       \log\left(\frac{\bar x}{x_i}\right) 
    =\log\left(\frac1n\sum_{i=1}^n x_i\right)
     -\frac1n\sum_{i=1}^n\log x_i.
\end{equation}
The Theil index is 
\begin{equation}
\label{eq:theil-definition-background}
    \operatorname{GE}_1(X) = T(X) 
    =\frac1n\sum_{i=1}^n
       \frac{x_i}{\bar x}
       \log\left(\frac{x_i}{\bar x}\right) 
    =\frac{1}{n\bar x}\sum_{i=1}^n x_i\log x_i
     -\log\bar x.
\end{equation}
These indices require all $x_i>0$. To ensure this in practice, organizations sometimes truncate or bottom-code income values. For example, the World Bank's Poverty and Inequality Platform replaces zero values with one
\citep{yonzan2026bottom}. When all incomes are strictly positive, both \eqref{eq:mld-definition} and \eqref{eq:theil-definition-background} are nonnegative. They equal zero if and only if all observations are identical.  They are increasingly large as the disparity   among the $x_i$ values grows.  

As a sense of the values seen in practice,  consider estimates from the UNU-WIDER World Income Inequality Database  Companion dataset in 2024. The MLD and Theil index of net per capita income for the United States are 0.32 and 0.31,  close to the respective medians of 0.30 and 0.31 across 210 countries' economies. Between the 5th and 95th percentiles, the MLD ranges from 0.14 to 0.68 and the Theil index from 0.13 to 0.66 
\citep{unuwider2026wiid}.

\subsection{Differential Privacy}
\label{subsec:dp-background}


For any two datasets $X$ and $X'$ each comprising $n$ individuals, 
define their Hamming distance as
\begin{equation}
\label{eq:hamming-distance}
    d(X,X')
    =\left|\{i\in\{1, \dots, n\}:x_i\ne x_i'\}\right|.
\end{equation}
We call $X$ and $X'$ neighboring databases when $d(X,X')=1$. 
A randomized mechanism $\mathcal M$ satisfies
$\varepsilon$-differential privacy if, for every pair of neighboring datasets $X$ and $X'$, and every measurable output set $\mathcal O$,
\begin{equation}
\label{eq:epsilon-dp}
    \Pr\{\mathcal M(X)\in\mathcal O\}
    \le e^\varepsilon\Pr\{\mathcal M(X')\in\mathcal O\}.
\end{equation}
The probabilities are over the  randomness of the
mechanism.  Intuitively, when $\epsilon$ is small, it is difficult to tell whether any particular output derives from $X$ or $X'$.  Hence, it is difficult to learn the value of the differing observation.  The literature on DP recommends values of $\epsilon\leq 1$, although in practice larger values are used to allow for greater accuracy \citep{kazan:reiter}.  

For $i=1, \dots, n$, suppose that  $L \leq x_i \leq U$ for all possible datasets $X \in [L, U]^n$ of size $n$.  Let $f$ be some statistic evaluated at $X$, which we write as $f(X)$. We define the local sensitivity of $f(X)$ as 
\begin{equation}
\label{eq:local-sensitivity-definition}
    \operatorname{LS}_f(X)
    =\max_{\substack{X'\in[L,U]^n\\d(X,X')=1}}
       |f(X')-f(X)|.
\end{equation}
The global sensitivity of $f$ is the maximum of the local sensitivities over all possible databases on size $n$, that is, 
\begin{equation}
\label{eq:global-sensitivity-definition}
    \operatorname{GS}_f
    =\max_{X\in[L,U]^n}\operatorname{LS}_f(X).
\end{equation}
Because the global sensitivity is determined by the most unfavorable datasets in the domain, it can substantially exceed the local sensitivity for the particular $X$ at hand.

The global sensitivity can be used to design differentially private algorithms for releasing a private version $\tilde{f}(X)$ of $f(X)$.  For example, one popular algorithm is the Laplace mechanism \citep{dwork2006calibrating}: add Laplace noise with scale $\operatorname{GS}_f/\varepsilon$ to $f(X)$, that is, $\tilde{f}(X) = f(X) + \eta$ where $\eta \sim Lap(0, \operatorname{GS}_f/\varepsilon)$. 
When the $\operatorname{GS}_f$ is reasonably small, the Laplace mechanism can yield $\tilde{f}(X)$ with low error.  However, when $\operatorname{GS}_f$ is large, it can result in high error. One would like to replace the 
global sensitivity with the generally smaller and dataset-specific local sensitivity; however, this does not generally preserve DP. The scale itself may change sharply between neighboring datasets. We therefore turn to smooth sensitivity, which 
allows the noise magnitude to depend on the observed data while ensuring the DP guarantee.

\subsection{Smooth Sensitivity and Private Release}
\label{subsec:smooth-background}

In reviewing smooth sensitivity, we closely follow the notation in  \citet{nissim2007smooth}. For  $k=0, \dots, n$, define
\begin{equation}
\label{eq:generic-Ak}
    A_f^{(k)}(X)
    =
    \max_{\substack{X'\in[L,U]^n\\d(X,X')\le k}}
       \operatorname{LS}_f(X').
\end{equation}
Here, 
$A_f^{(0)}(X)=\operatorname{LS}_f(X)$, and 
$A_f^{(n)}(X)=\operatorname{GS}_f$.
To compute  $A_f^{(k)}(X)$ for other values of $k$ with $X \in [L, U]^n$, one could proceed as follows: (i) replace some subset of $k$ observations in $X$ with $k$ other values in $[L, U]$ to form a candidate $X'$, (ii) compute the $\operatorname{LS}_f(X')$ for this candidate $X'$, and (iii) repeat these two steps for all possible replacement values in $[L, U]^k$ and all possible subsets of $k$ observations in $X$.  Naturally, this can be computationally expensive, thereby necessitating theory or approximations that circumvent the need to search over all possible datasets.  In Section \ref{sec:dp-mld} and Section \ref{sec:dp-theil}, we develop such theory and approximations for $M(X)$ and $T(X)$.

After determining the set of $\{A^{(k)}_f(X): k=0, \dots, n\}$, the analyst selects a smoothing parameter $\beta>0$ to define  the 
$\beta$-smooth sensitivity,
\begin{equation}
\label{eq:generic-smooth-sensitivity}
    \mathcal S_{f,\beta}(X)
    =
    \max_{X'\in[L,U]^n}
    e^{-\beta d(X,X')}
    \operatorname{LS}_f(X')
    =
    \max_{0\le k\le n}
    e^{-\beta k}A_f^{(k)}(X).
\end{equation}
More generally, a function
$\overline{\mathcal S}_{f,\beta}$ is a
$\beta$-smooth upper bound on the local sensitivity if
$\overline{\mathcal S}_{f,\beta}(X) \ge \operatorname{LS}_f(X)$ for every $X$, and $\overline{\mathcal S}_{f,\beta}(X) \le e^\beta
\overline{\mathcal S}_{f,\beta}(X')$ whenever $d(X,X')=1$. The exact smooth sensitivity is the
smallest such upper bound. A $\beta$-smooth upper bound can be used directly to construct a DP release. Let $Z_\gamma$ have density 
\begin{equation}
\label{eq:generic-poly-density}
    f(z) \propto \frac{1}{1+|z|^\gamma},
    \qquad \gamma>1.
\end{equation}
For a privacy budget $\varepsilon_f>0$, set
\begin{equation}
\label{eq:generic-noise-calibration}
    \alpha_f=\frac{\varepsilon_f}{4\gamma},
    \qquad
    \beta_f=\frac{\varepsilon_f}{\gamma},
\end{equation}
and release
\begin{equation}
\label{eq:generic-private-release}
    \widetilde f(X)
    =
    f(X)
    +
    \frac{
        \overline{\mathcal S}_{f,\beta_f}(X)
    }{\alpha_f}
    Z_\gamma.
\end{equation}
This mechanism satisfies $\varepsilon_f$-differential privacy.
In what follows, we set $\gamma=2$ so that $Z_\gamma$ is a standard Cauchy random variable, $Z\sim\operatorname{Cauchy}(0,1)$.



\section{Differential Privacy for the Mean Log Deviation}
\label{sec:dp-mld}

We now present the methodology for releasing a DP version of the MLD.    Section \ref{subsec:mld-local-global} first presents expressions for the local and global sensitivities of the MLD.   Section \ref{subsec:mld-Ak-main} then  characterizes the maximum local sensitivity for any dataset that differs in $1 \leq k \leq n$ data points from $X$; these are the values of $A^{(k)}_f(X)$ referenced in Section \ref{subsec:smooth-background}. Section \ref{subsec:mld-computation-main} discusses the computational complexity of the algorithm. We summarize the procedure for computing any $A^{(k)}_M(X)$ in  Algorithm~\ref{alg:lsb-mld} at the end of this section. 
We illustrate the computations on a toy dataset in Appendix \ref{sec:toy}.

\subsection{Local and Global Sensitivity of the MLD}
\label{subsec:mld-local-global}

For each $m = 1, \dots, n$,  let $S_m(X)= \sum_{i=1}^n x_i -x_m$.  We   
write $S_m=S_m(X)$ when the dataset is clear. When we replace  $x_m$ by some  $y\in[L,U]$, we denote the resulting dataset by
   $X^{(x_m\to y)}
    =(x_1,\ldots,x_{m-1},y,x_{m+1},\ldots,x_n).$
Then, using \eqref{eq:mld-definition}, the change in the MLD is 
\begin{equation}
\label{eq:mld-replacement-main}
    M\bigl(X^{(x_m\to y)}\bigr)-M(X)
    =
    \log\left(\frac{S_m+y}{S_m+x_m}\right)
    -\frac1n\log\left(\frac{y}{x_m}\right).
\end{equation}
For any $S\in[(n-1)L,(n-1)U]$ representing any value of the sum of $n-1$ elements, define
\begin{equation}
\label{eq:h-main}
    h_M(t;S)=\log(S+t)-\frac1n\log t,
    \qquad t\in[L,U].
\end{equation}
Then, we can write 
    $M\bigl(X^{(x_m\to y)}\bigr)-M(X)
    =h_M(y;S_m)-h_M(x_m;S_m).$ 

By taking the derivative of the function $h_M(t;S)$ with respect to $t$, we can show that,  for fixed $S$, $h_M( \cdot;S)$ decreases to a unique
minimum at $t_M^\star(S)=S/(n-1)\in[L,U]$ and increases thereafter. Consequently, for any $x\in[L,U]$ and for fixed sum $S$, the quantity representing the maximum absolute change in the MLD when replacing that $x$ with $y$, which we write as 
\begin{equation}
\label{eq:mld-D-definition}
    D_M(S,x)
    =\max_{y\in[L,U]}|h_M(y;S)-h_M(x;S)|,
\end{equation}
has the representation,
\begin{equation}
\label{eq:D-main}
    D_M(S,x)
    =\max\left\{
    \begin{aligned}
        &|h_M(L;S)-h_M(x;S)|,\\
        &|h_M(U;S)-h_M(x;S)|,\\
        &h_M(x;S)-h_M(t_M^\star(S);S)
    \end{aligned}
    \right\}.
\end{equation}

With this result, we state the local sensitivity of $M(X)$ in Proposition \ref{prop:mld-local-sensitivity}.
\begin{proposition}[Local sensitivity of the MLD]
\label{prop:mld-local-sensitivity}
For any $X\in[L,U]^n$,
\begin{equation}
\label{eq:mld-LS-main}
    \operatorname{LS}_M(X)
    =\max_{m\in \{1, \dots, n\}}D_M(S_m,x_m).
\end{equation}
\end{proposition}
\noindent The proof is given in
Appendix~\ref{app:mld-local-sensitivity}.
We  state the global sensitivity in Proposition \ref{prop:mld-global-sensitivity}.
\begin{proposition}[Global sensitivity of the MLD]
\label{prop:mld-global-sensitivity}
When neighboring datasets are in $[L,U]^n$,
\begin{equation}
\label{eq:mld-global-sensitivity-LU}
    \operatorname{GS}_M
    =
    \log\left(\frac{(n-1)L+U}{nL}\right)
    -\frac1n\log\left(\frac UL\right).
\end{equation}
The maximum is attained by the neighboring datasets
$(L,\ldots,L)$ and $(U,L,\ldots,L)$.
For $n=2$, the pair $(U,U)$ and $(L,U)$ attains the same maximum.
\end{proposition}
\noindent The proof is given in Appendix~\ref{app:mld-global-sensitivity}.
Unlike the local sensitivity, $\operatorname{GS}_M$ depends only on
$n, L,$ and $U$. 

\subsection{Smooth Sensitivity of the MLD}
\label{subsec:mld-Ak-main}

For each $0\leq k\leq n$, we seek to determine the quantity needed for smooth sensitivity of the MLD, 
\begin{equation}
\label{eq:Ak-main}
    A_M^{(k)}(X)
    =\max_{\substack{X'\in[L,U]^n\\d(X,X')\le k}}
        \operatorname{LS}_M(X').
\end{equation}
To do so, we define several terms that enter into the expression for \eqref{eq:Ak-main}.

First, consider any fixed $m\in \{1, \dots, n\}$. Let
$x_{m,[1]}\le\cdots\le x_{m,[n-1]}$ denote the ordered observations
excluding $x_m$. For $q=0,\ldots,n-1$, define the bounds 
\begin{align}
\label{eq:lower-S-main}
    \underline S_m^{(q)}(X)
    &=S_m-\sum_{j=1}^q\bigl(x_{m,[n-j]}-L\bigr),\\
\label{eq:upper-S-main}
    \overline S_m^{(q)}(X)
    &=S_m+\sum_{j=1}^q\bigl(U-x_{m,[j]}\bigr),
\end{align}
where an empty sum is zero. 
These are the smallest and largest attainable sums after excluding $x_m$
and modifying at most $q$ of the data points in $X$. 
Note that $\underline S_m^{(0)}(X)=\overline S_m^{(0)}(X)=S_m$. For convenience, we suppress the argument $X$ in these bounds when there is no ambiguity. Next, define
\begin{equation}
    R_M(S)=\max_{x\in[L,U]}D_M(S,x)
          =\max_{x,y\in[L,U]}\bigl|h_M(y;S)-h_M(x;S)\bigr|.\label{eq:RM}
\end{equation}
For any fixed  $S$ based on $n-1$ data points, \eqref{eq:RM} is the maximum change in the MLD over all possible values of the final data point.  It equals the range of $h_M(\cdot;S)$ on $[L,U]$, that is, its maximum minus its minimum. Since $h_M(\cdot;S)$ decreases on $[L,t_M^\star(S)]$ and increases on $[t_M^\star(S),U]$, its minimum is attained at $t_M^\star(S)\in[L,U]$ and its maximum at one of the endpoints $L$ or $U$. Therefore,
\begin{equation}
\label{eq:R-explicit-main}
    R_M(S)
    =\max\left\{
        h_M(L;S)-h_M(t_M^\star(S);S),
        h_M(U;S)-h_M(t_M^\star(S);S)
    \right\}.
\end{equation}

With these definitions established, we now  state the values of \eqref{eq:Ak-main} in Theorem \ref{thm:mld-Ak-main}.
\begin{theorem}[Exact local sensitivity in a $k$-neighborhood]
\label{thm:mld-Ak-main}
For $k=1,\ldots,n-1$,
\begin{equation}
\label{eq:Ak-exact-main}
A_M^{(k)}(X)
=
\max_{m\in \{1, \dots, n\}}
\max\left\{
D_M(\underline S_m^{(k)},x_m),
D_M(\overline S_m^{(k)},x_m),
R_M(\underline S_m^{(k-1)}),
R_M(\overline S_m^{(k-1)})
\right\}.
\end{equation}
For $k=0$ and $k=n$,
\begin{align}
    A_M^{(0)}(X)&=\max_{m\in \{1, \dots, n\}}D_M(S_m,x_m),
    \label{eq:A0-main}\\
    A_M^{(n)}(X)&=\operatorname{GS}_M.
    \label{eq:An-main}
\end{align}
\end{theorem}
The proof is given in Appendix~\ref{app:mld-smooth-proof}.  Theorem~\ref{thm:mld-Ak-main} reduces optimization over an entire $k$-neighborhood to four candidates for each $m$, which represents far fewer cases to check than an exhaustive search over all possible values in $[L,U]$. 

The exact $\beta$-smooth sensitivity is
\begin{equation}
\label{eq:mld-smooth-main}
    \mathcal S_{M,\beta}(X)
    =\max_{X'\in[L,U]^n}
        e^{-\beta d(X,X')}\operatorname{LS}_M(X')
    =\max_{k\ge0}e^{-\beta k}A_M^{(k)}(X).
\end{equation}

\begin{corollary}[Exact smooth sensitivity of the MLD]
\label{cor:mld-smooth}

For every $\beta>0$,
\begin{equation}
\label{eq:mld-smooth-exact-main}
    \mathcal S_{M,\beta}(X)
    =
    \max\left\{
        A_M^{(0)}(X),\;
        \max_{1\le k\le n-1}
        e^{-\beta k}A_M^{(k)}(X),\;
        e^{-\beta n}\operatorname{GS}_M
    \right\}.
\end{equation}
\end{corollary}
This follows from the 
representation of smooth sensitivity in \eqref{eq:generic-smooth-sensitivity} together with Theorem~\ref{thm:mld-Ak-main}. In particular,
$\mathcal S_{M,\beta}$ is the exact $\beta$-smooth upper bound on $\operatorname{LS}_M$.
Applying the generic mechanism in Section~\ref{subsec:smooth-background} with $f(X)=M(X)$, and using  $\alpha_M=\varepsilon_M/8$ and $\beta_M=\varepsilon_M/2$, we  release
\begin{equation}
\label{eq:private-mld-main}
    \widetilde M(X)
    =
    M(X)
    +
    \frac{
        {\mathcal S}_{M,\beta_M}(X)
    }{\alpha_M}Z_M,
    \qquad
    Z_M\sim\operatorname{Cauchy}(0,1).
\end{equation}

\begin{algorithm}[t]
\caption{$\textsc{SensBound}_M(X,k)$: compute any $A^{(k)}_M(X)$ for the MLD}
\label{alg:lsb-mld}
\begin{algorithmic}[1]
\If{$k=0$}
  \State \Return $\mathrm{LS}_M(X)$: the largest $|\Delta M|$ when some $x_m$ moves to $L$, $U$ or the average of the others
\EndIf
\For{$m=1,\dots,n$}
  \State Form four candidates by changing $k$ incomes
  \Statex \hspace{\algorithmicindent}(i) lower the $k$ largest of the other incomes to $L$
  \Statex \hspace{\algorithmicindent}(ii) raise the $k$ smallest of the other incomes to $U$
  \Statex \hspace{\algorithmicindent}(iii) same as (i), but change only $k-1$ of the other incomes, and use the last change to move $x_m$ to the average of the others
  \Statex \hspace{\algorithmicindent}(iv) same as (ii), but change only $k-1$ of the other incomes, and use the last change to move $x_m$ to the average of the others
  \State For each candidate, move $x_m$ to $L$, $U$ or the average of the others, and record $|\Delta M|$
\EndFor
\State \Return the largest recorded $|\Delta M|$
\end{algorithmic}
\end{algorithm}

\subsection{Computation}
\label{subsec:mld-computation-main}

The exact smooth sensitivity of the MLD can be computed in $O(n^2)$ time. We sort the observations once and form prefix sums. Then, for any $m \in \{1, \dots, n\}$ and for any  $q$, the sums of the
$q$ smallest and $q$ largest observations excluding $x_m$ can be obtained in constant time, which gives $\underline S_m^{(q)}$ and $\overline S_m^{(q)}$ in constant time. 
For each $k=1,\ldots,n-1$, we evaluate the four candidates in \eqref{eq:Ak-exact-main} for every $m = 1, \dots, n$ and take the
maximum. 
Since there are
$O(n^2)$ pairs $(m,k)$ and each pair requires only constant work after sorting, the overall algorithm runs in $O(n^2)$ time.

The search over $k$ can be  terminated early. If the largest weighted value found through  $q=k-1$ is at least $e^{-\beta k}\operatorname{GS}_M$, then no  $q>k-1$ 
can improve the current maximum, since
$A_M^{(\ell)}(X)\le\operatorname{GS}_M$ for all $\ell$. Thus, the computation may stop without changing the exact smooth sensitivity.

\section{Differential Privacy for the Theil Index}
\label{sec:dp-theil}

In this section, we present the algorithm for releasing a differentially private version of the Theil index, $T(X)$.  We again leverage smooth sensitivity to avoid potentially large noise from DP mechanisms that rely on global sensitivity.  However, unlike the MLD, for the Theil index we do not determine an exact expression for the maximum change in the local sensitivity when replacing $k$ data points in $X$; this turns out to be computationally impractical.  Rather, we construct a computable smooth upper bound on these maximum changes, which we then use to compute smooth sensitivity.

In Section \ref{subsec:theil-local-main}, we derive the local and global sensitivities of the Theil index. In Section \ref{subsec:theil-neighborhood-main}, we derive the computable smooth upper bound. In Section \ref{subsec:theil-computation-main}, we discuss the computational run time of the algorithm.  Algorithm~\ref{alg:lsb-theil}  summarizes the procedure for computing the smooth bounds on any $A^{(k)}_T(X)$ at the end of this section. 
The computations are illustrated in Appendix \ref{sec:toy-theil}.

\subsection{Local and Global Sensitivity}
\label{subsec:theil-local-main}

To begin, we define  notation used to derive the local and global sensitivity.  
For $m = 1, \dots, n$, let $S_m(X)=\sum_{i=1}^n x_i-x_m$ as before and 
\begin{equation}
    H_m(X)=\sum_{i=1}^n x_i\log x_i-x_m\log x_m.
\end{equation}
We suppress the argument $X$ when the dataset is clear.  For any
$S>0$ and $H\in\mathbb R$, define
\begin{equation}
\label{eq:theil-h-main}
    h_T(t;S,H)
    =\frac{H+t\log t}{S+t}-\log(S+t),
    \qquad t\in[L,U].
\end{equation}
Note that \eqref{eq:theil-h-main} is equivalent to the expression for the Theil index from \eqref{eq:theil-definition-background}. Using the notation of Section~\ref{sec:dp-mld}, the change in $T(X)$ when swapping out some $x_m \in X$ for some value $y \in [L, U]$ is
\begin{equation}
\label{eq:theil-replacement-main}
    T\bigl(X^{(x_m\to y)}\bigr)-T(X)
    =h_T(y;S_m,H_m)-h_T(x_m;S_m,H_m).
\end{equation}
For any $(S, H)$ jointly satisfying $(n-1)L\le S\le(n-1)U$ and $S\log L\le H\le S\log U$, we can take the derivative of $h_T(t;S,H)$ with respect to $t$ to show the unique minimum of $h_T(\cdot;S,H)$ occurs at
\begin{equation}
\label{eq:theil-minimizer-main}
    t_T^\star(S,H)=\exp(H/S)\in[L,U].
\end{equation}
Finally, we also define the maximum change in the Theil index at $(S, H)$ when changing $x$ to $y$ as
\begin{equation}
\label{eq:theil-D-definition-main}
    D_T(S,H,x)
    =\max_{y\in[L,U]}|h_T(y;S,H)-h_T(x;S,H)|.
\end{equation}
Since $\partial h_T(t;S,H)/\partial t=(S\log t-H)/(S+t)^2$, the function $h_T(\cdot;S,H)$ decreases on $[L,t_T^\star(S,H)]$ and increases on $[t_T^\star(S,H),U]$, so it attains its minimum at $t_T^\star(S,H)$ and its maximum at $L$ or $U$. Hence,
\begin{equation}
\label{eq:theil-D-main}
    D_T(S,H,x)
    =\max\left\{
    \begin{aligned}
        &|h_T(L;S,H)-h_T(x;S,H)|,\\
        &|h_T(U;S,H)-h_T(x;S,H)|,\\
        &h_T(x;S,H)-h_T(t_T^\star(S,H);S,H)
    \end{aligned}
    \right\}.
\end{equation}

With this notation, we state an expression for the local sensitivity of $T(X)$ in Proposition \ref{prop:theil-local-sensitivity}.
\begin{proposition}[Local sensitivity]
\label{prop:theil-local-sensitivity}
For every $X\in[L,U]^n$,
\begin{equation}
\label{eq:theil-LS-main}
    \operatorname{LS}_T(X)
    =\max_{m\in\{1, \dots, n\}}D_T(S_m,H_m,x_m).
\end{equation}
\end{proposition}
\noindent The proof is given in Appendix~\ref{app:theil-local-sensitivity}. We state the global sensitivity in Proposition \ref{prop:theil-global-sensitivity}.
\begin{proposition}[Global sensitivity]
\label{prop:theil-global-sensitivity}
When neighboring databases are in $[L,U]^n$,
\begin{equation}
\label{eq:theil-global-sensitivity-LU}
    \operatorname{GS}_T
    =
    \frac{U}{(n-1)L+U}
    \log\left(\frac UL\right)
    -
    \log\left(
        \frac{(n-1)L+U}{nL}
    \right).
\end{equation}
The maximum is attained by the neighboring datasets $(L,\ldots,L)$ and $(U,L,\ldots,L)$. For $n=2$, the pair
$(U,U)$ and $(L,U)$ attains the same maximum.
\end{proposition}
\noindent The proof is given in Appendix~\ref{app:theil-global-sensitivity}. 

\subsection{Bounds and Smooth Sensitivity for the Theil Index}
\label{subsec:theil-neighborhood-main}

For $0 \leq k \leq n$, we define the quantities relevant for smooth sensitivity as
\begin{equation}
\label{eq:theil-Ak-main}
    A_T^{(k)}(X)
    =
    \max_{\substack{X'\in[L,U]^n\\d(X,X')\le k}}
    \operatorname{LS}_T(X').
\end{equation}
We find and use an upper bound on each  $A_T^{(k)}(X)$.  We use  upper bounds because 
the full set of pairs $(S_m,H_m)$ attainable after changing $k$ income values is not computable sufficiently fast for a viable algorithm.


For any feasible $(S,H)$, let $R_T(S,H)=
    \max_{x\in[L,U]}D_T(S,H,x)$  be the maximum possible change in the Theil index when changing any element $x$ to some $y$ with the remaining elements summing to $S$ and $H$. This change occurs when moving $x$ from $t_T^\star$, which offers the minimum value in the U-shaped curve  $h_T(\cdot;S,H)$, to one of the endpoints $L$ or $U$. Based on \eqref{eq:theil-D-main}, we have 
\begin{equation}
\label{eq:theil-R-main}
    R_T(S,H)
    =
    \max_{c\in\{L,U\}}
    \left\{
        h_T(c;S,H)
        -
        h_T(t_T^\star(S,H);S,H)
    \right\}.
\end{equation}
Since $R_T(S,H)$ itself is a maximum, we must have $A_T^{(k)}(X)$ no larger than the largest among these maxima. That is,  
\begin{equation}
\label{eq:theil-Ak-relaxation-main}
    A_T^{(k)}(X)
    \le
    \max_{\substack{X'\in[L,U]^n,\ d(X,X')\le k\\m\in \{1, \dots, n\}}}
    R_T(S_m(X'),H_m(X')).
\end{equation}

While \eqref{eq:theil-Ak-relaxation-main} suggests an upper bound on $A_T^{(k)}(X)$, it is cumbersome to compute over all possible $X'$ and $m$.  We therefore derive bounds on the possible values of any $(S_m, H_m)$ obtained by replacing $k$ values of $X$ with other plausible values in $[L,U]^k$. Let $x_{(1)}\le\cdots\le x_{(n)}$ be the ordered observations.
Define $\phi(x)=x\log x$ for any value of $x$. Let
$v_{(1)}\le\cdots\le v_{(n)}$ be the ordered values of
$\{v_i=\phi(x_i): x_i \in X\}$. Since $\phi$ is convex, 
\begin{eqnarray}
\label{eq:theil-phi-extrema-main}
    \phi_{\min}
    &=&
    \begin{cases}
        -e^{-1},&L\le e^{-1}\le U,\\
        \min\{L\log L,U\log U\},&\text{otherwise}.
    \end{cases}\\
    \phi_{\max}
    &=&
    \max\{L\log L,U\log U\}.
\end{eqnarray}
Using this notation, we can establish lower and upper bounds on any $S$ and $H$ when replacing  $q_k=\min\{k,n-1\}$ elements of $X$. We have
\begin{align}
    \underline S^{(k)}(X)
    &=\sum_{i=1}^{n-q_k-1}x_{(i)}+q_kL,\label{eq:theil-S-lower-main}\\
    \overline S^{(k)}(X)
    &=\sum_{i=q_k+2}^{n}x_{(i)}+q_kU,\label{eq:theil-S-upper-main}\\
    \underline H^{(k)}(X)
    &=\sum_{i=1}^{n-q_k-1}v_{(i)}+q_k\phi_{\min},\label{eq:theil-H-lower-main}\\
    \overline H^{(k)}(X)
    &=\sum_{i=q_k+2}^{n}v_{(i)}+q_k\phi_{\max}.\label{eq:theil-H-upper-main}
\end{align}
Lemma~\ref{lem:theil-summary-extrema} in Appendix~\ref{app:theil-technical-lemmas} shows that these are the exact marginal extrema of $S_m(X')$ and $H_m(X')$ over all $m$ and all $X'$ with $d(X,X')\le k$. Each bound is attained, although generally not jointly.
Hence, every feasible pair of $(S, H)$ is an element of the set defined by these bounds, 
\begin{equation}
\label{eq:theil-rectangle-main}
    \mathcal C^{(k)}(X)
    =
    [\underline S^{(k)}(X),\overline S^{(k)}(X)]
    \times
    [\underline H^{(k)}(X),\overline H^{(k)}(X)].
\end{equation}
Because $S>0$, the minimum and maximum of $H/S$ over
$\mathcal C^{(k)}(X)$ occur at its corners. Define
\begin{align}
\label{eq:theil-eta-lower-main}
    \underline\eta^{(k)}(X)
    &=
    \min_{(S,H)\in\mathcal C^{(k)}(X)}
    \frac{H}{S},
\\
\label{eq:theil-eta-upper-main}
    \overline\eta^{(k)}(X)
    &=
    \max_{(S,H)\in\mathcal C^{(k)}(X)}
    \frac{H}{S},
\end{align}
where the extrema are obtained by checking the four corner ratios.
Set
\begin{align}
\label{eq:theil-t-lower-main}
    \underline t_T^{(k)}(X)
    &=
    \max\{L,\exp(\underline\eta^{(k)}(X))\},
\\
\label{eq:theil-t-upper-main}
    \overline t_T^{(k)}(X)
    &=
    \min\{U,\exp(\overline\eta^{(k)}(X))\}.
\end{align}
Then, every feasible $X'$ and $m$ satisfy
\begin{equation}
\label{eq:theil-t-containment-main}
    \underline t_T^{(k)}(X)
    \le
    t_T^\star(S_m(X'),H_m(X'))
    \le
    \overline t_T^{(k)}(X).
\end{equation}
To bound the two endpoint terms in \eqref{eq:theil-R-main}, 
for $c\in\{L,U\}$ we define
\begin{equation}
\label{eq:theil-endpoint-function-main}
    f_{T,c}(S,t)
    =
    \frac{c}{S+c}\log\left(\frac ct\right)
    +
    \log\left(\frac{S+t}{S+c}\right).
\end{equation}
When $t=t_T^\star(S,H)$, we have
$
    h_T(c;S,H)-h_T(t;S,H)
    =
    f_{T,c}(S,t).
$
Moreover, since
\begin{equation}
    \frac{\partial}{\partial t}f_{T,c}(S,t)
    =
    \frac{S(t-c)}{t(S+c)(S+t)}, 
\end{equation}
$f_{T,L}(S,t)$ is nondecreasing in $t\in[L,U]$ and $f_{T,U}(S,t)$ is nonincreasing.  Thus, define
\begin{align}
\label{eq:theil-BL-main}
    B_{T,L}^{(k)}(X)
    &=
    \max_{S\in[\underline S^{(k)}(X),\overline S^{(k)}(X)]}
    f_{T,L}(S,\overline t_T^{(k)}(X)),
\\
\label{eq:theil-BU-main}
    B_{T,U}^{(k)}(X)
    &=
    \max_{S\in[\underline S^{(k)}(X),\overline S^{(k)}(X)]}
    f_{T,U}(S,\underline t_T^{(k)}(X)),
\end{align}
and let
\begin{equation}
\label{eq:theil-Braw-main}
    \widetilde B_T^{(k)}(X)
    =
    \max\left\{
        B_{T,L}^{(k)}(X),
        B_{T,U}^{(k)}(X)
    \right\}.
\end{equation}
Finally, using the global sensitivity as an additional bound, we arrive at the upper bounds on each $A_T^{(k)}(X)$,
\begin{equation}
\label{eq:theil-Bk-main}
    B_T^{(k)}(X)
    =
    \min\left\{
        \operatorname{GS}_T,
        \widetilde B_T^{(k)}(X)
    \right\}.
\end{equation}
This result is stated formally in 
Theorem \ref{thm:theil-neighborhood-bound}. 
\begin{theorem}[A neighborhood upper bound for the Theil index]
\label{thm:theil-neighborhood-bound}
For every $X\in[L,U]^n$ and every integer $k\ge0$,
$   A_T^{(k)}(X)
    \le
    B_T^{(k)}(X).$
\end{theorem}
\noindent The proof is given in 
Appendix~\ref{app:theil-neighborhood-proof}.
We use the bounds $B_T^{(k)}(X)$ from Theorem~\ref{thm:theil-neighborhood-bound} to define 
\begin{equation}
\label{eq:theil-smooth-bound-main}
    \overline{\mathcal S}_{T,\beta}(X)
    =
    \max_{0\le k\le n-1}
    e^{-\beta k}B_T^{(k)}(X).
\end{equation}
The search stops at $n-1$ because 
both $\widetilde B_T^{(k)}$ and $B_T^{(k)}$  are constant
for $k\ge n-1$. Thus, we can state Theorem \ref{thm:theil-smooth-bound} providing the smooth upper bound for the Theil index.

\begin{theorem}[A Smooth upper bound for the Theil index]
\label{thm:theil-smooth-bound}
For every $\beta>0$, $\overline{\mathcal S}_{T,\beta}$ is a $\beta$-smooth upper bound on the local sensitivity of the Theil index. In particular,
\begin{equation}
\label{eq:theil-smooth-domination-main}
    \overline{\mathcal S}_{T,\beta}(X)
    \ge
    \mathcal S_{T,\beta}(X)
    \ge
    \operatorname{LS}_T(X),
\end{equation}
and, whenever $d(X,Y)=1$,
\begin{equation}
\label{eq:theil-smoothness-main}
    \overline{\mathcal S}_{T,\beta}(X)
    \le
    e^\beta
    \overline{\mathcal S}_{T,\beta}(Y).
\end{equation}
\end{theorem}
\noindent The proof is given in Appendix~\ref{app:theil-smooth-proof}.


Applying the generic smooth-sensitivity mechanism in
Section~\ref{subsec:smooth-background}, for a privacy budget
$\varepsilon_T>0$ set $\alpha_T=\varepsilon_T/8$ and $\beta_T=\varepsilon_T/2$, we release
\begin{equation}
\label{eq:private-theil-main}
    \widetilde T(X)
    =
    T(X)
    +
    \frac{
        \overline{\mathcal S}_{T,\beta_T}(X)
    }{\alpha_T}Z_T,
    \qquad
    Z_T\sim\operatorname{Cauchy}(0,1).
\end{equation}

\subsection{Computational Run Time}
\label{subsec:theil-computation-main}

For fixed $X$ and $k$, the definitions of $B_{T,L}^{(k)}(X)$ and $B_{T,U}^{(k)}(X)$ already fix the second argument of $f_{T,c}(S,t)$, i.e., $(c,t) = (L,\overline t_T^{(k)})$ and $(U,\underline t_T^{(k)})$, respectively. It remains to maximize each function over $S\in[\underline S^{(k)},\overline S^{(k)}]$. We evaluate these maximizations directly, without a numerical search.

For either choice of $(c,t)$, if $t=c$, then
$f_{T,c}(S,t)=0$ for every $S>0$. Otherwise, let $r=t/c$.
Differentiating with respect to $S$ gives
\begin{equation}
    \frac{\partial}{\partial S}f_{T,c}(S,t)
    =
    \frac{
        c\{c(r\log r-r+1)-(r-1-\log r)S\}
    }
    {(S+c)^2(S+t)}.
\end{equation}
For $r>0$ with $r\ne1$, both
$r-1-\log r$ and $r\log r-r+1$ are positive. Thus, the derivative
changes from positive to negative at
\begin{equation}
    S^\star
    =
    c\frac{r\log r-r+1}{r-1-\log r}.
\end{equation}
Consequently, $f_{T,c}(S,t)$ increases up to $S^\star$ and
decreases thereafter.

On $[\underline S^{(k)},\overline S^{(k)}]$, the maximizing
value of $S$ is $S^\star$ if it lies in the interval,
$\underline S^{(k)}$ if $S^\star$ lies below it, and
$\overline S^{(k)}$ if $S^\star$ lies above it.
Evaluating $f_{T,c}(S,t)$ at this value gives the corresponding
endpoint bound. Once the summary bounds are available, both
$B_{T,L}^{(k)}(X)$ and $B_{T,U}^{(k)}(X)$ therefore take
constant time to compute. Taking their maximum and applying the
global-sensitivity cap in \eqref{eq:theil-Bk-main} gives
$B_T^{(k)}(X)$ in constant time as well.

For the overall computation, we have to sort $x_i$ and $\phi_i$ and form cumulative sums, taking $O(n\log n)$ time.
The four summary bounds and the corresponding bounds on
$t_T^\star$ can be obtained in constant time per $k$.
Evaluating $B_T^{(k)}(X)$ for $k=0,\ldots,n-1$ and taking
the maximum of $e^{-\beta k}B_T^{(k)}(X)$ requires only
$O(n)$ additional work. Hence,
$\overline{\mathcal S}_{T,\beta}(X)$ is computed in
$O(n\log n)$ total time and $O(n)$ storage.

\begin{algorithm}[t]
\caption{$\textsc{SensBound}_T(X,k)$: compute upper bound $B^{(k)}_T(X)\ge A^{(k)}_T(X)$ for the Theil index}
\label{alg:lsb-theil}
\begin{algorithmic}[1]
\State $[\underline S,\overline S]\gets$ smallest and largest total of $n-1$ incomes after $k$ changes: lower the $k$ largest to $L$ and exclude the next largest; raise the $k$ smallest to $U$ and exclude the next smallest
\State $[\underline H,\overline H]\gets$ smallest and largest total of  $x_i\log x_i$ for $n-1$ incomes after $k$ changes: lower the $k$ largest to $L$ and exclude the next largest; raise the $k$ smallest to $U$ and exclude the next smallest (when $L\ge e^{-1}$, e.g.\ $L=1$)
\State $[\underline t,\overline t]\gets$ range of $\exp(H/S)$ over $[\underline S,\overline S]\times[\underline H,\overline H]$, truncated to $[L,U]$
\State $B_U\gets\max_{S\in[\underline S,\overline S]} f_{T,U}(S,\underline t)$
\State $B_L\gets\max_{S\in[\underline S,\overline S]} f_{T,L}(S,\overline t)$
\Statex \Comment{both maximizations in closed form, Sec.~\ref{subsec:theil-computation-main}}
\State \Return $\min\bigl\{\mathrm{GS}_T,\ \max(B_U,B_L)\bigr\}$
\end{algorithmic}
\end{algorithm}

\section{Simulation Studies}
\label{sec:experiments}

We evaluate the accuracy of the differentially private MLD and Theil algorithms using two sets of simulation studies.  In Section \ref{subsec:realdata}, we run each algorithm repeatedly on household-level and individual-level income data from the CPS. Here, we measure the differences in the noisy indices relative to the values of the indices in the CPS data.  In Section \ref{sec:simulation}, we perform repeated sampling from a population and compare the noisy indices to the simulated population-level index values. In both simulation studies, we use  $f\in\{M,T\}$ to refer to each index. 

\subsection{Application to CPS Income Data}
\label{subsec:realdata}


We use the 2025 CPS Annual Social and Economic Supplement  public-use files \citep{census2025asec}, which report income for 2024. The personal income sample contains $107{,}969$ adults aged 18 or older with positive person weights and nonmissing, nonnegative total income (\texttt{PTOTVAL}). The
household income sample contains $55{,}695$ households with at least one member, an identified householder, a positive household weight, and nonmissing, nonnegative total income (\texttt{HTOTVAL}).  We compute the indices without using the survey weights.

We evaluate both indices using  $x_i=\mathrm{income}_i+1$ to avoid zeros, and we set the lower bound $L=1$. As the upper bound $U$ on incomes, we do not use a very large value known to exceed any possible income.  Doing so would result in large variances in the noise distributions. Instead, we use a differentially private algorithm to estimate a private upper bound $\widetilde U(X)$ on the data values; see  Appendix \ref{sec:private-upper-bound} for details.  We allocate an additional privacy budget of $\varepsilon_U=0.15$ to compute this noisy upper bound. Because it is generated stochastically, it is possible for $\widetilde U(X)$ to be less than some of the largest observed incomes.  When this occurs, we  
clip the data to the noisy bound, so that we input  $X^c=\{x_i^c: i=1,\dots,n\}$ with $x_i^c=\min\{x_i,\widetilde U(X)\}$ into the algorithms for computing the noisy estimates.  For each $f(X^{c})$, we use a privacy 
budget of $\varepsilon_f \in\{0.25,0.5,1,1.5,2\}$, which is varied to assess the impacts of the privacy budget on accuracy. 

For each sample and index, we generate $r=1, \dots, 10{,}000$ releases at each $\varepsilon_f$. Each release $r$  uses the same $\widetilde U^{(r)}$ for all $\varepsilon_f$, though we generate a new $\widetilde  U^{(r)}$ for each $r$. 
We also store the number of values that are clipped at the upper bound and whether $\widetilde U^{(r)}<\max_i x_i$.



\begin{figure}[t]
\centering
\includegraphics[width=\textwidth]{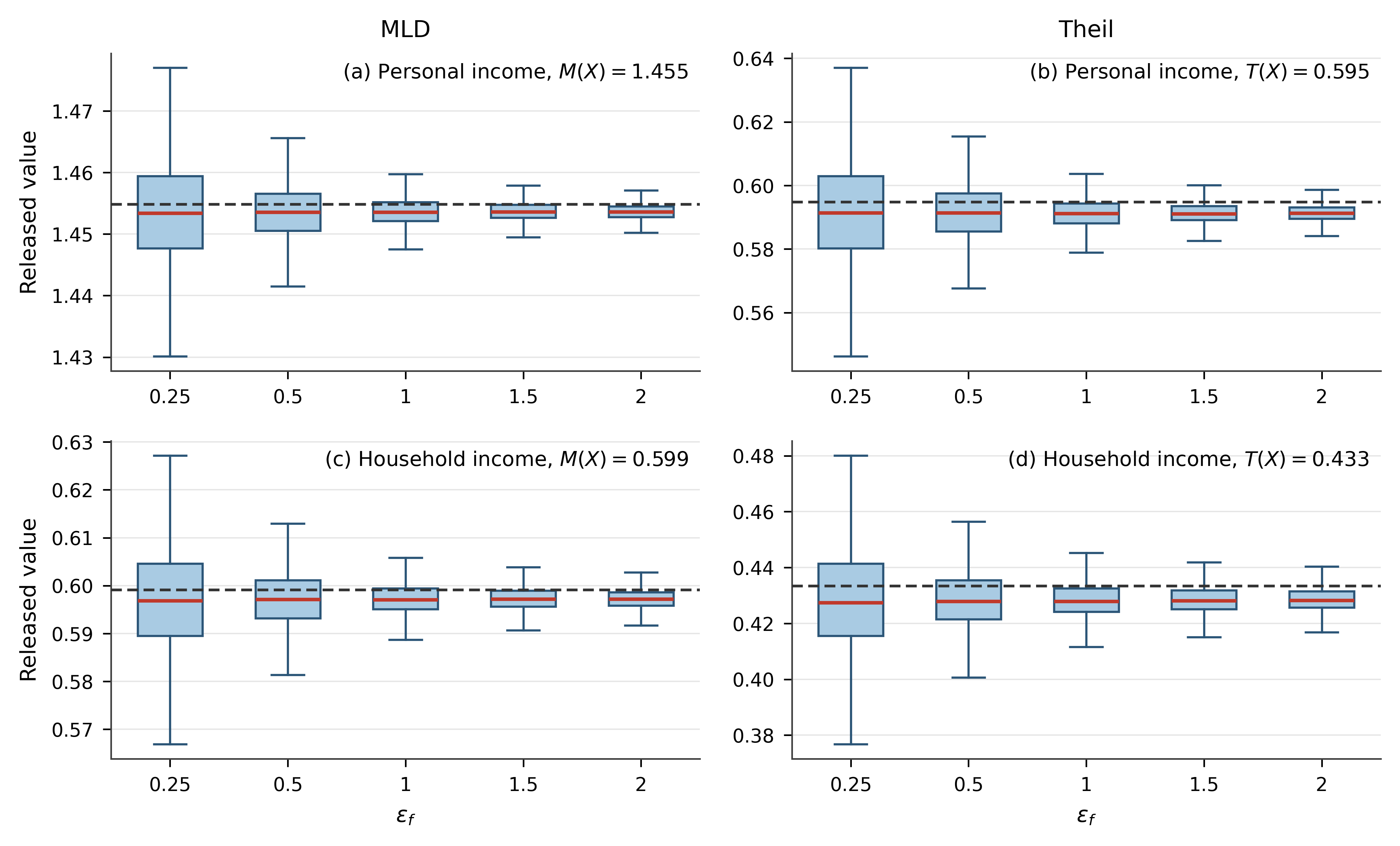}
\caption{Distributions of the 10,000 simulated values of the differentially private MLD and Theil indices across different privacy budgets. Panels (a) and (b) show the MLD and Theil index for personal income. Panels (c) and (d) show the indices for household income. The dashed lines represent the values of the indices computed without any privacy adjustments.}
\label{fig:release-boxplots}
\end{figure}

Figure~\ref{fig:release-boxplots} 
displays the distributions of the 10,000 randomly drawn values of the differentially private indices for different values of $\varepsilon_f$.  The differentially private estimates typically are close to the confidential values. As expected, as $\varepsilon_f$ increases, the 
absolute errors tend to decrease.  The small but persistent negative bias results because the algorithm clips observations in runs where $\widetilde U^{(r)}<\max_i x_i$. In fact, with $\varepsilon_U=0.15$, some form of clipping occurs in $97.25\%$ of the simulation runs for personal income releases and $95.49\%$ of the simulation runs for household income releases. However, the median numbers of clipped records are only $17$ and $24$, which represent $0.016\%$ and $0.043\%$ of the sample sizes, respectively. While the clipping introduces bias, it also lowers the value of 
the smooth sensitivity, which in turn shrinks the scale of the Cauchy noise. In fact, at $\varepsilon_f=1$, the releases based on $\widetilde U(X)$ have smaller median and 95th-percentile absolute errors than (non-private) implementations of the algorithms that set $U=\max_i x_i$, for both indices and both samples, even though the former spends an additional $\varepsilon_U=0.15$ on selecting the bound.
Hence, we generally 
suggest using $\widetilde U(X)$ in the differentially private algorithm for computing the MLD or Theil index.

\subsection{Repeated Sampling Study}
\label{sec:simulation}

We next examine the performance of the differentially private indices in repeated sampling studies.
We generate incomes from a zero-inflated GB2 (generalized Beta of the second kind) distribution calibrated to approximately match the empirical distribution of the personal-income CPS data.
 Specifically, we generate income values from the density 
\begin{equation}
\label{eq:simulation-zero-gb2}
    f(y) = \pi_0\delta_0+(1-\pi_0)\operatorname{GB2}(a,b,p,q),
\end{equation}
where $\delta_0$ is a point mass at zero and the GB2 component has density
\begin{equation}
\label{eq:simulation-gb2-density}
    g_{\mathrm{GB2}}(y;a,b,p,q)
    =\frac{a}{b\,B(p,q)}\left(\frac{y}{b}\right)^{ap-1}
      \left\{1+\left(\frac{y}{b}\right)^a\right\}^{-(p+q)},
    \qquad y>0,
\end{equation}
with $a,b,p,q>0$ and $B(\cdot,\cdot)$ the beta function.   We set $\pi_0=0.084$ to match the  proportion of zero incomes in the CPS data. We set $(a,b,p,q)$ by maximizing the survey-weighted log-likelihood $\sum_{i:y_i>0}w_i\log g_{\mathrm{GB2}}(y_i;a,b,p,q)$ over the positive incomes $y_i$ in the CPS personal income sample, where $w_i$ is the CPS
person weight. This gives
    $(a,b,p,q)
    =(4.169,\;79605.35,\;0.194,\;0.492)$.
For purposes of computing the MLD and Theil index, we set $x_i=y_i+1$ for all $i$.   
We use numerical integration to determine the expected MLD and Theil index under the data generation mechanism. These values are   $M(F)\approx1.465$ and  $T(F)\approx0.685$, respectively. We note that these are reasonably close to the indices for the individual incomes in the CPS data.


We consider sample sizes $n\in\{10^3,10^4,10^5,10^6\}$ and 
$\varepsilon_f\in\{0.5,1,2\}$, and we run $R=1,000$ independent replications at each $n$. We use a lower bound on the simulated incomes of $L=1$. We use a differentially private upper bound 
as in Section~\ref{subsec:realdata} with
$\varepsilon_U=0.15$. In each replication $r$, we simulate a sample $X^{(r)}$ of size $n$ from the data generation mechanism. In each $r$, we use the same $\widetilde U(X^{(r)})$ when computing the smooth sensitivity across the three values of $\varepsilon_f$.

Let $\widetilde f(X^{c(r)})$ be the
private release and $f(X^{(r)})$ be the nonprivate index in replication $r$. We measure their errors,
\begin{equation}
\label{eq:simulation-errors}
    E_f^{(r)}=|\widetilde f(X^{c(r)})-f(F)|,
    \qquad
    E_{\mathrm{NP},f}^{(r)}=|f(X^{(r)})-f(F)|.
\end{equation}
The first expression is the total error in the private release. The second expression is the error introduced by random sampling of observed data, which provides a benchmark error measure. 


\begin{figure}[t]
\centering
\includegraphics[width=0.85\textwidth]{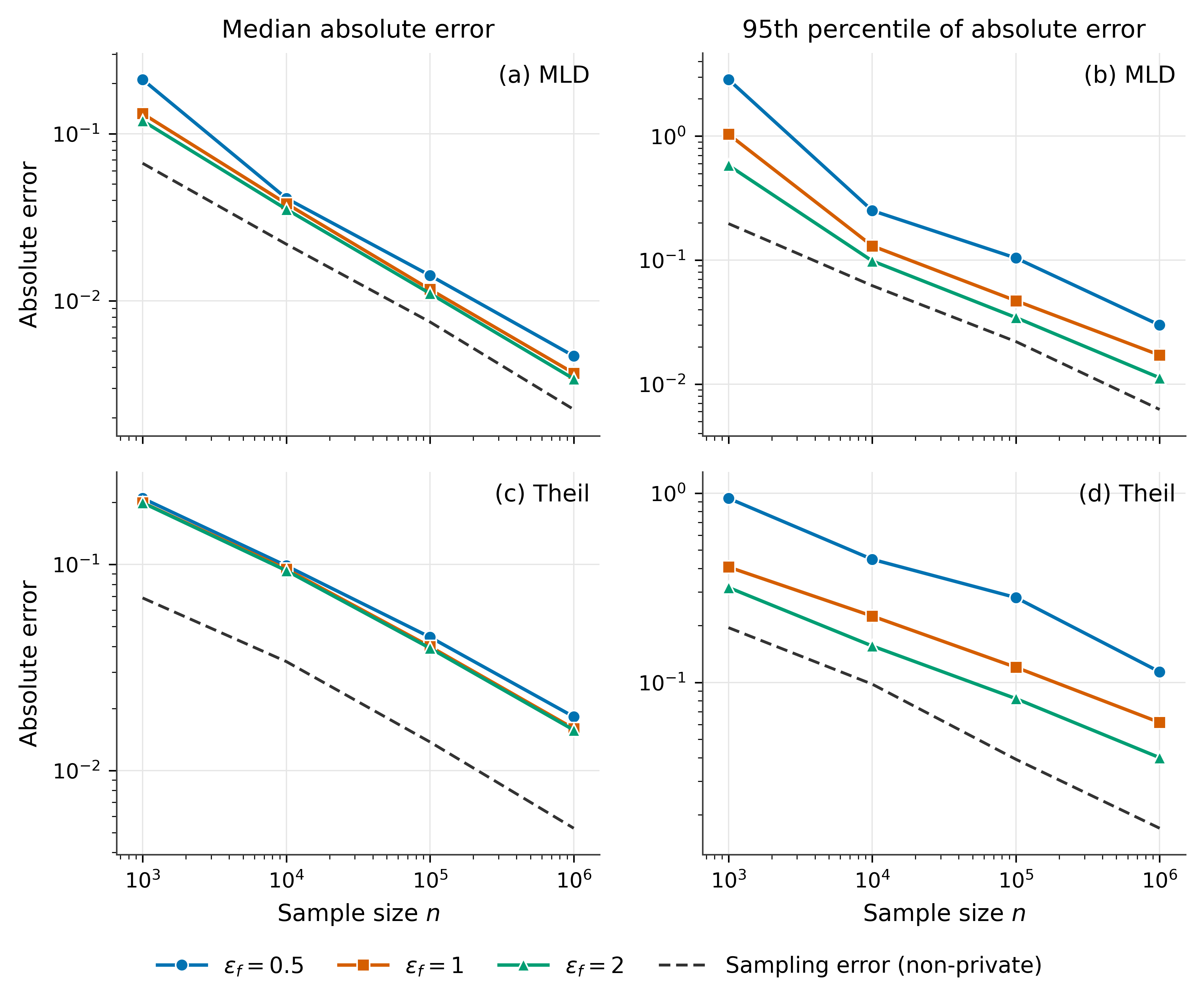}
\caption{Differences in the differentially private indices versus their corresponding values derived from the data generation model, plotted against sample size on log--log axes.  Panels (a)--(b) show the MLD and panels (c)--(d)
the Theil index; the left column shows the medians and the right column the 95th percentiles of the absolute errors. The dashed line represents the absolute errors of the non-private values of the indices, which arise purely from random sampling from the data generating model.}
\label{fig:simulation-errors}
\end{figure}


Figure~\ref{fig:simulation-errors} 
displays the median and 95th percentile of the total errors $E_f^{(r)}$ and 
the sampling errors $E_{\mathrm{NP},f}^{(r)}$ for both indices. 
Several broad trends are apparent.  First, as expected, the absolute errors tend to 
decrease as  $n$ and $\varepsilon_f$ increase. 
Second, for equivalent sample sizes with $n>1,000$, the absolute errors tend to be smaller for the MLD than the Theil index. The Theil index is more sensitive than the MLD to values in the upper tails, and it is those values that drive its smooth sensitivity. Since the GB2 distribution has a heavy upper tail, truncating some large incomes at $\widetilde U(X^{(r)})$ biases the index downward. 
Third, differences in $\varepsilon_f$ mainly affect the tail behavior of the algorithms. For example, at $n=10,000$, increasing $\varepsilon_f$ from $0.5$ to $2$ reduces the 95th-percentile error from $0.251$ to $0.099$ for the MLD and from $0.447$ to $0.156$ for the Theil index, whereas the medians change only from $0.041$ to $0.035$ and from $0.098$ to $0.093$.  Fourth, when $n=1,000$, the typical errors can be quite large, especially for the Theil index.  Arguably, at these privacy budgets, the noise introduced to ensure differential privacy is too high to yield reliable estimates at small sample sizes.
Finally, for the MLD in particular, the typical error in these scenarios is small (other than when $n=1,000$).  Larger $n$ is needed to ensure small absolute errors at the 95th percentile. For the Theil index, the typical errors are larger though still reasonable when $n>1,000$; however, the 95th percentile  of absolute errors remains large at $\varepsilon_f=0.5$ for most of these sample sizes.

\section{Discussion}
\label{sec:discussion}

We develop differentially private algorithms for the MLD and the Theil index. Both take advantage of the fact that
the effect of changing any one income on the index depends 
on summaries of the other incomes. For the MLD, this structure is simple enough that we 
can  compute the smooth sensitivity exactly. For the Theil index, the two totals cannot be controlled jointly in closed form; thus, we bound them separately.  In the CPS and artificial data  simulations, the MLD releases have small errors at moderate budgets. The Theil releases tend to have larger errors, though the algorithm is practicable  for larger sample sizes and budgets.
It may be possible to reduce the errors 
by finding a tighter smooth upper bound that captures how the two 
totals occur jointly. 

There are several directions for future research. First, one could develop differentially private versions of other members of the generalized entropy family.  The strategy of bounding summary statistics that we used for the differentially private Theil index may be useful for other indices.
Second, it should be possible to extend the algorithms to sub-groups. One appealing feature of generalized entropy indices is their additive decomposability. Releasing sub-group statistics and the aggregated measure in a differentially private manner raises questions of how to allocate the budget across groups and how to handle small groups. Third, we did not use  survey weights in the indices.  
Differential privacy with survey weights is difficult. Changing one respondent can change both their income and their weight, and when weights are calibrated to population totals it changes the weights of other respondents as well \citep{reiterarisa, drechsler:dp, seeman}. Extending the algorithms to survey-weighted statistics may require redefining neighboring datasets and deriving the sensitivities under that definition.

\bibliography{reference}

\newpage

\begin{appendices}

\section{Illustrative Examples of the Computations}
In this section, we illustrate the computations for each differentially private algorithm using a small, exemplary dataset. Throughout this section, we take the bounds $L$ and $U$ as given. 

\subsection{Example of the MLD}
\label{sec:toy}
 
We illustrate the algorithm for the MLD with $n=5$ incomes, $X=(2,\,3,\,5,\,8,\,12)$. Here, $M(X)=0.1987$. We set $(L, U) =(1,20)$ and a budget $\varepsilon_M=1$, so that $\alpha_M=\varepsilon_M/8=0.125$ and $\beta_M=\varepsilon_M/2=0.5$.  By Proposition~\ref{prop:mld-global-sensitivity}, the global sensitivity on $[1,20]^5$ is $\mathrm{GS}_M=0.9695$, about five times $M(X)$. It is attained by changing $(1,1,1,1,1)$ into $(20,1,1,1,1)$.

\paragraph{Step 1: the local sensitivity.}
To find $\mathrm{LS}_M(X)$, we change one income at a time and record the resulting change $|\Delta M|=|M(X')-M(X)|$, where $X'$ is the modified dataset. 
Proposition~\ref{prop:mld-local-sensitivity} shows that when individual $m$ moves to a new value $y$, the other four individuals affect the change in $M(X)$ only through their total income. As a function of $y$, the MLD is U-shaped, with its minimum where $y$ equals the average income of the other four. Hence only three values of $y$ need to be checked: $L$, $U$, and that average.
 
For example, take $x_3=5$. The other four incomes sum to $25$, so their average is $6.25$. The three candidate neighbors are
\[
\begin{aligned}
(2,3,\mathbf{1},8,12):&\quad M(X')=0.3774,\quad |\Delta M|=0.1788,\\
(2,3,\mathbf{20},8,12):&\quad M(X')=0.3269,\quad |\Delta M|=0.1282,\\
(2,3,\mathbf{6.25},8,12):&\quad M(X')=0.1948,\quad |\Delta M|=0.0038.
\end{aligned}
\]
Because $x_3=5$ is already close to the minimum, moving it there changes $M(X)$ little; moving it to $L$ changes $M(X)$ the most. Repeating this for each $x_i$ gives $\mathrm{LS}_M(X)=0.1788$, attained by moving $x_3$ to $L$. The local sensitivity  is less than one-fifth of the $\mathrm{GS}_M$.
 
\paragraph{Step 2: the local sensitivity after changing $k$ incomes.}
Smooth sensitivity  requires the values of $A^{(k)}_M(X)$, the largest value of the local sensitivity of any dataset obtained by changing up to $k$ incomes in $X$. A brute-force search over a $100$-point grid of $[1,20]$ would examine $\binom{5}{k}100^k$ datasets for each $k$. As this number suggests, a brute force search would be infeasible for datasets with large $n$ and hence large $k$.  Theorem~\ref{thm:mld-Ak-main} dramatically reduces this search, as we now illustrate. 

Suppose the incomes are in increasing order (as is $X$).  For each individual $m=1, \dots, n$, the algorithm determines which
dataset $X'$, where $d(X, X') \leq k$, results in the largest value of $|M(X') - M(X'')|$, where $X''$ is constructed by replacing the $m$th element of $X'$ with some other value in $[L,U]$.   
To construct $X'$, the $k$ changes can be distributed in one of two ways. First, suppose all $k$ changes are to incomes other than individual $m$.  By Step~1, these changes matter only through the total income among these $k$ other individuals. The largest  $|M(X') - M(X'')|$ occurs when that total is pushed to an extreme: either the $k$ largest of the other incomes are lowered to $L$, or the $k$ smallest of the other incomes are raised to $U$. Second, and alternatively, suppose one change is to $x_m$ and $k-1$ changes are to other incomes.  The largest  $|M(X') - M(X'')|$ occurs when pushing the total of the $k-1$ other incomes to one of the two extremes, i.e., lowering the $k-1$ largest to $L$ or raising the $k-1$ smallest to $U$, and moving $x_m$ to the average income of the remaining individuals.

We now provide the computations for $k=2$ and $m=3$, showing the four candidates ($X'$) when we consider changing $x_3=5$ to compute  $|M(X') - M(X'')|$, as well as the largest value of that change in the MLD. The four candidates include
\[
\begin{aligned}
&\text{lower } 8,12 \text{ to } L: && (2,3,5,\mathbf{1},\mathbf{1})\ \to\ (2,3,\mathbf{20},1,1), && |\Delta M|=0.5337,\\
&\text{raise } 2,3 \text{ to } U: && (\mathbf{20},\mathbf{20},5,8,12)\ \to\ (20,20,\mathbf{1},8,12), && |\Delta M|=0.2584,\\
&\text{lower } 12 \text{ to } L, \text{ move } x_3 \text{ to } 3.5: && (2,3,\mathbf{3.5},8,\mathbf{1})\ \to\ (2,3,\mathbf{20},8,1), && |\Delta M|=0.3156,\\
&\text{raise } 2 \text{ to } U, \text{ move } x_3 \text{ to } 10.75: && (\mathbf{20},3,\mathbf{10.75},8,12)\ \to\ (20,3,\mathbf{1},8,12), && |\Delta M|=0.2748.
\end{aligned}
\]
We repeat this process over all five individuals, searching for the largest change in the MLD.  In this case, it happens to occur for $m=3$, resulting in   $A^{(2)}_M(X)=0.5337$.

Table~\ref{tab:toy-mld} lists, for each $k$, a dataset $X'$ such that $d(X, X') \leq k$ with the largest local sensitivity and the change that attains it. In every row the worst case follows the same pattern: for $k\ge1$, the $k$ largest incomes are lowered to $L$, and the largest remaining income is then raised to $U$. As $k$ grows, $X'$ moves toward $(1,1,1,1,1)$, from which the change to $(20,1,1,1,1)$ attains $\mathrm{GS}_M$. This pattern arises because the incomes in $X$ lie close to $L$ relative to $U$. For  a set of incomes close to $U$, such as $(15,17,18,19,20)$, the worst case usually instead raises the smallest incomes to $U$ and lowers a remaining income to $L$. Theorem~\ref{thm:mld-Ak-main} therefore checks all four candidates.
 
\begin{table}[t]
\centering
\begin{tabular}{@{}c l c r r@{}}
\toprule
$k$ & $X'$ & final change & $A^{(k)}_M(X)$ & $e^{-\beta_M k}A^{(k)}_M(X)$ \\
\midrule
0 & $(2,3,5,8,12)$ & $5\to1$  & 0.1788 & 0.1788 \\
1 & $(2,3,5,8,\mathbf1)$  & $8\to20$ & 0.3063 & 0.1858 \\
2 & $(2,3,5,\mathbf1,\mathbf1)$  & $5\to20$ & 0.5337 & \textbf{0.1963} \\
3 & $(2,3,\mathbf1,\mathbf1,\mathbf1)$  & $3\to20$ & 0.7600 & 0.1696 \\
4 & $(2,\mathbf1,\mathbf1,\mathbf1,\mathbf1)$  & $2\to20$ & 0.9258 & 0.1253 \\
5 & $(\mathbf1,\mathbf1,\mathbf1,\mathbf1,\mathbf1)$  & $1\to20$ & 0.9695 & 0.0796 \\
\bottomrule
\end{tabular}
\caption{Step 2 for the MLD example. For each $k$, $X'$ is a dataset within distance $k$ of $X$ that attains $A^{(k)}_M(X)$. The column labeled ``final change'' attains the maximum change in the MLD at that $k$.}
\label{tab:toy-mld}
\end{table}
 
\paragraph{Step 3: smoothing and release.}
The smooth sensitivity is the largest weighted value in Table~\ref{tab:toy-mld}, attained at $k=2$:
\[
\mathcal S_{M,\beta_M}(X)=0.1963,
\qquad
\widetilde M=0.1987+\frac{0.1963}{0.125}\,Z_M=0.1987+1.571\,Z_M,
\qquad Z_M\sim\mathrm{Cauchy}(0,1).
\]

\subsection{Example of the Theil Index}
\label{sec:toy-theil}

In this section, we illustrate the computations for the differentially private Theil index algorithm.  We use the same dataset and budget as in Appendix \ref{sec:toy}. Here, $T(X)=0.1810$. By  Proposition~\ref{prop:theil-global-sensitivity}, $\mathrm{GS}_T=0.9278$, where the global sensitivity is attained by changing $(1,1,1,1,1)$ into $(20,1,1,1,1)$.

\paragraph{Step 1: the local sensitivity.}
When individual $m$ moves from income $x_m$ to a new value $y$, the other four individuals' incomes affect the change in $T(X)$ through two totals: their total income $S_m$ and the total $H_m=\sum_{i\ne m}x_i\log x_i$, as stated in  Proposition~\ref{prop:theil-local-sensitivity}. As a function of $y$, the Theil index is  U-shaped. Its minimum occurs at $t^\star_T=\exp(H_m/S_m)$. Hence, to find the maximal amount $T(X)$ can change when altering $x_m$, we only need to consider $y \in \{L, U, t^\star_T\}$.   

For example, take $x_4=8$. The other four incomes give $S_4=22$, $H_4=42.548$ and $t^\star_T=6.917$. The three candidate neighbors include 
\[
\begin{aligned}
(2,3,5,\mathbf{1},12):&\quad |\Delta T|=0.1428,\\
(2,3,5,\mathbf{20},12):&\quad |\Delta T|=0.1303,\\
(2,3,5,\mathbf{6.917},12):&\quad |\Delta T|=0.0020.
\end{aligned}
\]
Repeating this for each individual in $X$ gives $\mathrm{LS}_T(X)=0.1428$, which happens to be attained by moving $x_4$ to $L$.

\paragraph{Step 2: bounding the local sensitivity after changing $k$ incomes.}
For the MLD, the other individuals' incomes are part of only one total, so the worst-case local sensitivity can be found by pushing that total to its extremes.  For the Theil index, these incomes are part of the pair of totals $(S_m,H_m)$, and the worst case depends on how the two totals move together. Theorem~\ref{thm:theil-neighborhood-bound} avoids accounting for this joint movement. It pushes each total to its extremes separately and bounds the change in $T(X)$ over every combination therein.

To illustrate, consider $k=1$ and change the income for some individual $m$.   
To find bounds on the resulting change in $T(X)$, we first find the smallest possible value of $T(X)$ when fixing $(S_m, H_m)$. 
Since $T(X)$ is U-shaped in any individual's income, its 
smallest value for a fixed $(S_m, H_m)$ occurs when the income for individual $m$ is at  
$t^\star_T=\exp(H_m/S_m)$.  The largest possible change in the Theil index occurs when changing this income from $t^\star_T$ to either $L$ or $U$. This change depends only on the other four incomes through $S_m$ and $H_m$. We therefore find how small and how large $S_m$ can be after one change. To make $S_m$ as small as possible, we lower the largest income $12$ to $L$ and leave out the next largest $8$. To make it as large as possible, we raise the smallest income $2$ to $U$ and leave out the next smallest $3$.  With these actions, we have 
\[
\begin{aligned}
&\text{exclude } 8,\ \text{lower } 12 \text{ to } L: && (2,3,5,\mathbf{1}), && S_m=11,\ H_m=12.73,\\
&\text{exclude } 3,\ \text{raise } 2 \text{ to } U: && (5,8,12,\mathbf{20}), && S_m=45,\ H_m=114.42.
\end{aligned}
\]
Since $x\log x$ increases with $x$ on $[1,20]$, these two datasets also provide the smallest and largest $H_m$. 

In the algorithm, we use the two ranges, $S_m\in[11,45]$ and $H_m\in[12.73,114.42]$. We do not determine which $S_m$ goes with which $H_m$. Thus, the algorithm allows $t^\star_T$ anywhere in $[1.327,20]$:
\[
\begin{aligned}
&S_m=45,\ H_m=12.73: && t^\star_T=\exp(12.73/45)=1.327,\\
&S_m=11,\ H_m=114.42: && t^\star_T=\exp(114.42/11),\ \text{capped at } U=20.
\end{aligned}
\]

The change in $T(X)$ when moving the income from $t^\star_T$ to $U$ is largest when $t^\star_T$ is as far from $U$ as possible, that is, $t^\star_T=1.327$. The change in $T(X)$ when moving the income to $L$ is largest when $t^\star_T=20$. For each, the worst-case value of $S_m\in[11,45]$ is found in closed form (Section~\ref{subsec:theil-computation-main}). Here, it is $S_m=11$. Thus, we have
\[
B^{(1)}_T=\max\bigl\{\underbrace{f_{T,U}(11,1.327)}_{\text{move to }U}=0.8280,\ \underbrace{f_{T,L}(11,20)}_{\text{move to }L}=0.6994\bigr\}=0.8280 .
\]
Because the algorithm allows pairs of $(S_m,H_m)$ that no dataset attains, $B^{(1)}_T$ is an upper bound on $A^{(1)}_T(X)$ rather than its exact value. 
Table~\ref{tab:toy-theil} repeats this calculation for each $k$.

\begin{table}[t]
\centering
\begin{tabular}{@{}c l l c r r@{}}
\toprule
$k$ & smallest $S$ & largest $S$ & range of $t^\star_T$ & $B^{(k)}_T$ & $e^{-\beta_T k}B^{(k)}_T$ \\
\midrule
0 & $(2,3,5,8)$                       & $(3,5,8,12)$                        & $[2.854,20]$ & 0.5355 & \textbf{0.5355} \\
1 & $(2,3,5,\mathbf1)$                & $(5,8,12,\mathbf{20})$              & $[1.327,20]$ & 0.8280 & 0.5022 \\
2 & $(2,3,\mathbf1,\mathbf1)$         & $(8,12,\mathbf{20},\mathbf{20})$    & $[1.081,20]$ & 0.9278 & 0.3413 \\
3 & $(2,\mathbf1,\mathbf1,\mathbf1)$  & $(12,\mathbf{20},\mathbf{20},\mathbf{20})$ & $[1.019,20]$ & 0.9278 & 0.2070 \\
4 & $(\mathbf1,\mathbf1,\mathbf1,\mathbf1)$ & $(\mathbf{20},\mathbf{20},\mathbf{20},\mathbf{20})$ & $[1,20]$ & 0.9278 & 0.1256 \\
\bottomrule
\end{tabular}
\caption{Step 2 for the Theil example. For each  $k$,  the second and third columns display the other four incomes at the smallest and largest attainable total of four income values. The range of $t^\star_T$ is computed from all four combinations of their $S$ and $H$, and $B^{(k)}_T$ is capped at $\mathrm{GS}_T$.}
\label{tab:toy-theil}
\end{table}

\paragraph{Step 3: smoothing and release.}
The smooth upper bound is the largest weighted value in Table~\ref{tab:toy-theil}, attained at $k=0$:
\[
\overline{\mathcal S}_{T,\beta_T}(X)=0.5355,
\qquad
\widetilde T=0.1810+\frac{0.5355}{0.125}\,Z_T=0.1810+4.284\,Z_T,
\qquad Z_T\sim\mathrm{Cauchy}(0,1).
\]

\section{Proofs for Smooth Sensitivity of  Mean Log Deviation}
\label{app:mld-proofs}

Throughout this section, we use the notation of Section~\ref{sec:dp-mld}. When the reference dataset is $X$, write $S_m=S_m(X)$ and suppress
$X$ in $\underline S_m^{(q)}(X)$ and $\overline S_m^{(q)}(X)$.

\subsection{Proof of Local Sensitivity}
\label{app:mld-local-sensitivity}

\renewcommand{\restatename}{Proposition~\ref{prop:mld-local-sensitivity}}
\begin{restate}[Local sensitivity of the MLD]
For any $X\in[L,U]^n$,
\begin{equation}
    \operatorname{LS}_M(X)
    =\max_{m\in \{1, \dots, n\}}D_M(S_m,x_m).
    \tag{\ref{eq:mld-LS-main}}
\end{equation}
\end{restate}

\begin{proof}
Fix $m\in \{1, \dots, n\}$. Replacing $x_m$ by $y\in[L,U]$ gives
\begin{align}
\label{eq:app-mld-replacement}
    M\bigl(X^{(x_m\to y)}\bigr)-M(X)
    &=\log\left(\frac{S_m+y}{S_m+x_m}\right)
      -\frac1n\log\left(\frac{y}{x_m}\right)\nonumber\\
    &=h_M(y;S_m)-h_M(x_m;S_m).
\end{align}
For $S\in[(n-1)L,(n-1)U]$, we have 
\begin{equation}
\label{eq:app-h-derivative}
    \frac{\partial}{\partial t}h_M(t;S)
    =\frac{1}{S+t}-\frac{1}{nt}
    =\frac{(n-1)t-S}{nt(S+t)}.
\end{equation}
Thus, $h_M(\cdot;S)$ attains its minimum at $t_M^\star(S)=S/(n-1)$ and
its maximum at one of $L$ and $U$. For any fixed $x\in[L,U]$, the largest absolute difference from
$h_M(x;S)$ must therefore occur at $L$, $U$, or $t_M^\star(S)$.
This gives \eqref{eq:D-main}, and hence
\begin{equation}
    \operatorname{LS}_M(X)
    =\max_{m\in\{1, \dots, n\}}\max_{y\in[L,U]}
       |h_M(y;S_m)-h_M(x_m;S_m)|
    =\max_{m\in\{1, \dots, n\}}D_M(S_m,x_m).
\end{equation}
\end{proof}

\subsection{Technical Lemmas}
\label{app:mld-reachable-sums}

Fix $m\in \{1, \dots, n\}$, and let
$x_{m,[1]}\le\cdots\le x_{m,[n-1]}$ be the ordered observations
excluding $x_m$. The bounds
$\underline S_m^{(q)}$ and $\overline S_m^{(q)}$ are described in 
\eqref{eq:lower-S-main} and \eqref{eq:upper-S-main}.

\begin{lemma}[Reachable range]
\label{lem:app-reachable-range}
Fix $m\in\{1, \dots, n\}$, and let $q\in\{0,\ldots,n-1\}$. If at most $q$
elements other than $x_m$ are modified, then
\begin{equation}
\label{eq:app-reachable-interval}
    \underline S_m^{(q)}
    \le S_m(X')
    \le \overline S_m^{(q)}.
\end{equation}
\end{lemma}

\begin{proof}
The smallest possible value of $S_m(X')$ is obtained by replacing
the $q$ largest observations among $\{x_i:i\ne m\}$ with $L$, giving
\begin{equation}
    \underline S_m^{(q)}
    =
    S_m-\sum_{j=1}^q
    \bigl(x_{m,[n-j]}-L\bigr).
\end{equation}
Similarly, the largest possible value is obtained by replacing the
$q$ smallest observations with $U$, giving
\begin{equation}
    \overline S_m^{(q)}
    =
    S_m+\sum_{j=1}^q
    \bigl(U-x_{m,[j]}\bigr).
\end{equation}
These two constructions use at most $q$ replacements, so both bounds
are attainable.
\end{proof}

\begin{lemma}[Endpoint reduction for fixed $x$]
\label{lem:app-D-endpoint}
Fix $x\in[L,U]$. For every interval
$[a,b]\subseteq[(n-1)L,(n-1)U]$,
\begin{equation}
\label{eq:app-D-endpoint}
    \max_{S\in[a,b]}D_M(S,x)
    =\max\{D_M(a,x),D_M(b,x)\}.
\end{equation}
\end{lemma}

\begin{proof}
Write $D_M(S,x) =\max\{d_{M,L}(S,x),d_{M,U}(S,x),d_{M,\star}(S,x)\}$, where $d_{M,L}(S,x)=|h_M(L;S)-h_M(x;S)|$, $d_{M,U}(S,x)=|h_M(U;S)-h_M(x;S)|$ and $d_{M,\star}(S,x)=h_M(x;S)-h_M(t_M^\star(S);S)$. We have
\begin{align*}
    \frac{\partial}{\partial S}
      \{h_M(L;S)-h_M(x;S)\}
    &=\frac{1}{S+L}-\frac{1}{S+x}\ge0,\\
    \frac{\partial}{\partial S}
      \{h_M(U;S)-h_M(x;S)\}
    &=\frac{1}{S+U}-\frac{1}{S+x}\le0.
\end{align*}
Each difference is monotone in $S$, so its absolute value cannot
exceed the larger absolute value at $a$ and $b$. For the remaining term, differentiation gives
\begin{equation}
\label{eq:app-dstar-derivative}
    \frac{\partial}{\partial S}d_{M,\star}(S,x)
    =\frac{1}{S+x}-\frac{n-1}{nS}
    =\frac{S-(n-1)x}{nS(S+x)}.
\end{equation}
Thus, $d_{M,\star}(S,x)$ is nonincreasing for $S\le(n-1)x$
and nondecreasing for $S\ge(n-1)x$. Its maximum on $[a,b]$
is therefore attained at an endpoint as well. Thus,
\begin{equation}
    D_M(S,x)\le\max\{D_M(a,x),D_M(b,x)\}.
\end{equation}
The reverse inequality for the maximum follows because both
endpoints belong to $[a,b]$.
\end{proof}

\begin{lemma}[Endpoint reduction for $R_M$]
\label{lem:app-R-endpoint}
For every interval $[a,b]\subseteq[(n-1)L,(n-1)U]$,
\begin{equation}
\label{eq:app-R-endpoint}
    \max_{S\in[a,b]}R_M(S)
    =\max\{R_M(a),R_M(b)\}.
\end{equation}
\end{lemma}

\begin{proof}
Recall the definition of $R_M$,
\begin{equation}
 R_M(S)
    =\max_{x,y\in[L,U]}|h_M(y;S)-h_M(x;S)|
    =\max\{g_{M,L}(S),g_{M,U}(S)\},
\end{equation}
where $g_{M,c}(S)=h_M(c;S)-h_M(t_M^\star(S);S)$ for $c\in\{L,U\}$. Also,
\begin{equation}
    g_{M,L}'(S)
    =\frac{S-(n-1)L}{nS(S+L)}\ge0, \qquad
    g_{M,U}'(S)
    =\frac{S-(n-1)U}{nS(S+U)}\le0.
\end{equation}
Hence, for every $S\in[a,b]$,
\begin{equation}
    g_{M,L}(S)\le g_{M,L}(b),
    \qquad
    g_{M,U}(S)\le g_{M,U}(a).
\end{equation}
It follows that
\begin{equation}
    R_M(S)
    \le\max\{g_{M,L}(b),g_{M,U}(a)\}
    \le\max\{R_M(a),R_M(b)\}.
\end{equation}
Since $a$ and $b$ are feasible, equality holds after maximizing
over $S\in[a,b]$.
\end{proof}
    
\subsection{Proof of Global Sensitivity}
\label{app:mld-global-sensitivity}

\renewcommand{\restatename}{Proposition~\ref{prop:mld-global-sensitivity}}
\begin{restate}[Global sensitivity of the MLD]
When neighboring datasets are in $[L,U]^n$,
\begin{equation}
    \operatorname{GS}_M
    =
    \log\left(\frac{(n-1)L+U}{nL}\right)
    -\frac1n\log\left(\frac UL\right).
    \tag{\ref{eq:mld-global-sensitivity-LU}}
\end{equation}
The maximum is attained by the neighboring datasets
$(L,\ldots,L)$ and $(U,L,\ldots,L)$.
For $n=2$, the pair $(U,U)$ and $(L,U)$ attains the same maximum.
\end{restate}

\begin{proof}
Proposition~\ref{prop:mld-local-sensitivity} gives
\begin{equation}
    \operatorname{GS}_M
    =\max_{X\in[L,U]^n}\max_{m\in\{1, \dots, n\}}D_M(S_m,x_m).
\end{equation}
For a fixed element $m$, any pair $(S,x)\in[(n-1)L,(n-1)U]\times[L,U]$ can be realized as $(S_m,x_m)$: set $x_m=x$ and all other observations equal to $S/(n-1)$. Therefore,
\begin{equation}
\operatorname{GS}_M=\max_{S\in[(n-1)L,(n-1)U]}
      \max_{x\in[L,U]}D_M(S,x)=\max_{S\in[(n-1)L,(n-1)U]}R_M(S).
\end{equation}
By Lemma~\ref{lem:app-R-endpoint},
\begin{equation}
\label{eq:app-GS-two-corners}
    \operatorname{GS}_M
    =\max\{R_M((n-1)L),R_M((n-1)U)\}.
\end{equation}
At $S=(n-1)L$, the minimum of $h_M(\cdot;S)$ occurs at $L$,
and hence
\begin{equation}
    R_M((n-1)L)
    =\log\left(\frac{(n-1)L+U}{nL}\right)
     -\frac1n\log\left(\frac UL\right).
\end{equation}
At $S=(n-1)U$, the minimum occurs at $U$, giving
\begin{equation}
    R_M((n-1)U)
    =\log\left(\frac{(n-1)U+L}{nU}\right)
     +\frac1n\log\left(\frac UL\right).
\end{equation}
Let $r=U/L>1$ and denote these two values by $\rho_A$ and
$\rho_B$, respectively. Then
\begin{equation}
\rho_A=\log\left(\frac{n-1+r}{n}\right)-\frac1n\log r,\qquad \rho_B=\log\left(\frac{n-1+1/r}{n}\right)+\frac1n\log r.
\end{equation}
Their difference is
\begin{equation}
\label{eq:app-q-r}
    q(r)=\rho_A-\rho_B
    =\log\left(\frac{n-1+r}{(n-1)r+1}\right)
      +\frac{n-2}{n}\log r.
\end{equation}
Since $q(1)=0$ and
\begin{equation}
\label{eq:app-q-derivative}
    q'(r)
    =\frac{(n-2)(n-1)(r-1)^2}
           {nr(n+r-1)((n-1)r+1)}\ge0,
\end{equation}
we have $\rho_A\ge\rho_B$. Equality holds for $n=2$; for
$n\ge3$ and $r>1$, the inequality is strict. Therefore,
\begin{equation}
    \operatorname{GS}_M
    =\rho_A
    =\log\left(\frac{(n-1)L+U}{nL}\right)
     -\frac1n\log\left(\frac UL\right).
\end{equation}
This value is attained by replacing one observation in
$(L,\ldots,L)$ with $U$. For $n=2$, replacing one observation
in $(U,U)$ with $L$ attains $\rho_B=\rho_A$ as well.
\end{proof}

\subsection{Proof of the Exact Smooth Sensitivity}
\label{app:mld-smooth-proof}

\renewcommand{\restatename}{Theorem~\ref{thm:mld-Ak-main}}
\begin{restate}[Exact local sensitivity in a $k$-neighborhood]
For $k=1,\ldots,n-1$,
\begin{equation}
A_M^{(k)}(X)
=
\max_{m\in \{1, \dots, n\}}
\max\left\{
D_M(\underline S_m^{(k)},x_m),
D_M(\overline S_m^{(k)},x_m),
R_M(\underline S_m^{(k-1)}),
R_M(\overline S_m^{(k-1)})
\right\}.
\tag{\ref{eq:Ak-exact-main}}
\end{equation}
For $k=0$ and $k=n$,
\begin{align}
    A_M^{(0)}(X)&=\max_{m\in \{1, \dots, n\}}D_M(S_m,x_m),
    \tag{\ref{eq:A0-main}}\\
    A_M^{(n)}(X)&=\operatorname{GS}_M.
    \tag{\ref{eq:An-main}}
\end{align}
\end{restate}

\begin{proof}
For any $X'=(x_1',\ldots,x_n')\in[L,U]^n$,
Proposition~\ref{prop:mld-local-sensitivity} implies
\begin{equation}
    \operatorname{LS}_M(X')
    =\max_{m\in\{1, \dots, n\}}D_M(S_m(X'),x_m').
\end{equation}
Consequently,
\begin{align}
\label{eq:app-Ak-interchange}
    A_M^{(k)}(X)
    &=\max_{\substack{X'\in[L,U]^n\\d(X,X')\le k}}
      \max_{m\in\{1, \dots, n\}}D_M(S_m(X'),x_m')\nonumber\\
    &=\max_{m\in\{1, \dots, n\}}
      \max_{\substack{X'\in[L,U]^n\\d(X,X')\le k}}
      D_M(S_m(X'),x_m').
\end{align}
When $k=0$, the dataset is just $X'=X$, which gives \eqref{eq:A0-main}. Now fix $k\in\{1,\ldots,n-1\}$ and $m\in\{1, \dots, n\}$. Define
\begin{align}
\label{eq:app-B0}
    \mathcal N_{m,0}^{(k)}(X)
    &=\{X'\in[L,U]^n:d(X,X')\le k,\ x_m'=x_m\},\\
\label{eq:app-B1}
    \mathcal N_{m,1}^{(k)}(X)
    &=\left\{X'\in[L,U]^n:
       \bigl|\{i\ne m:x_i'\ne x_i\}\bigr|\le k-1\right\}.
\end{align}
In the second set, $x_m'$ ranges freely over $[L,U]$.
Both sets lie in the $k$-neighborhood of $X$, and
\begin{equation}
\label{eq:app-ball-decomposition}
    \{X'\in[L,U]^n:d(X,X')\le k\}
    =\mathcal N_{m,0}^{(k)}(X)
     \cup\mathcal N_{m,1}^{(k)}(X).
\end{equation}
On $\mathcal N_{m,0}^{(k)}(X)$, element $m$ is fixed.
Lemma~\ref{lem:app-reachable-range} shows that $S_m(X')$ ranges
exactly over $[\underline S_m^{(k)},\overline S_m^{(k)}]$.
Using Lemma~\ref{lem:app-D-endpoint},
\begin{align}
\label{eq:app-case0}
    &\max_{X'\in\mathcal N_{m,0}^{(k)}(X)}
       D_M(S_m(X'),x_m')\nonumber
    =\max_{S\in[\underline S_m^{(k)},\overline S_m^{(k)}]}
       D_M(S,x_m)\nonumber\\
    &\quad=\max\{
       D_M(\underline S_m^{(k)},x_m),
       D_M(\overline S_m^{(k)},x_m)\}.
\end{align}
On $\mathcal N_{m,1}^{(k)}(X)$, the sum of the other elements
ranges exactly over
$[\underline S_m^{(k-1)},\overline S_m^{(k-1)}]$.
For every attainable sum, $x_m'$ can be chosen independently
anywhere in $[L,U]$, with at most $k$ total changes. Thus
Lemma~\ref{lem:app-R-endpoint} gives
\begin{align}
\label{eq:app-case1}
    &\max_{X'\in\mathcal N_{m,1}^{(k)}(X)}
       D_M(S_m(X'),x_m')\nonumber
    =\max_{S\in[\underline S_m^{(k-1)},\overline S_m^{(k-1)}]}
       \max_{x\in[L,U]}D_M(S,x)\nonumber\\
    &\quad=\max_{S\in[\underline S_m^{(k-1)},\overline S_m^{(k-1)}]}
       R_M(S)\nonumber\\
    &\quad=\max\{
       R_M(\underline S_m^{(k-1)}),
       R_M(\overline S_m^{(k-1)})\}.
\end{align}
Combining \eqref{eq:app-ball-decomposition},
\eqref{eq:app-case0}, and \eqref{eq:app-case1}, and then maximizing over $m$, proves \eqref{eq:Ak-exact-main}.

Finally, every dataset in $[L,U]^n$ is within Hamming distance
$n$ of $X$. Therefore,
\begin{equation}
    A_M^{(n)}(X)
    =\max_{X'\in[L,U]^n}\operatorname{LS}_M(X')
    =\operatorname{GS}_M,
\end{equation}
which proves \eqref{eq:An-main}.
\end{proof}

\section{Proofs for Smooth Sensitivity Bound of Theil Index}
\label{app:theil-proofs}

Throughout this appendix, we use the notation of Section~\ref{sec:dp-theil}.

\subsection{Proof of Local Sensitivity}
\label{app:theil-local-sensitivity}

\renewcommand{\restatename}{Proposition~\ref{prop:theil-local-sensitivity}}
\begin{restate}[Local sensitivity]
For every $X\in[L,U]^n$,
\begin{equation}
    \operatorname{LS}_T(X)
    =\max_{m\in\{1, \dots, n\}}D_T(S_m,H_m,x_m).
    \tag{\ref{eq:theil-LS-main}}
\end{equation}
\end{restate}

\begin{proof}
As stated in \eqref{eq:theil-replacement-main}, replacing $x_m$ with
$y\in[L,U]$ changes the Theil index by
\begin{equation}
    T\bigl(X^{(x_m\to y)}\bigr)-T(X)
    =
    h_T(y;S_m,H_m)-h_T(x_m;S_m,H_m).
\end{equation}
For fixed $(S,H)$, 
\begin{equation}
    \frac{\partial}{\partial t}h_T(t;S,H)
    =
    \frac{S\log t-H}{(S+t)^2}.
\end{equation}
For every $(S_m,H_m)$,
\begin{equation}
    \frac{H_m}{S_m}
    =
    \sum_{i\ne m}\frac{x_i}{S_m}\log x_i
    \in[\log L,\log U],
\end{equation}
since the weights $x_i/S_m$ are nonnegative and sum to one.
Hence,
\begin{equation}
    t_T^\star(S_m,H_m)
    =
    \exp(H_m/S_m)
    \in[L,U].
\end{equation}
The derivative is negative below $t_T^\star$ and positive above it,
so $h_T(\cdot;S_m,H_m)$ is minimized at $t_T^\star$ and maximized
at one of the endpoints $L$ or $U$. Therefore,
\begin{equation}
    \max_{y\in[L,U]}
    \left|
        h_T(y;S_m,H_m)-h_T(x_m;S_m,H_m)
    \right|
    =
    D_T(S_m,H_m,x_m).
\end{equation}
Maximizing over $m$ gives
\begin{equation}
    \operatorname{LS}_T(X)
    =
    \max_{m\in\{1, \dots, n\}}D_T(S_m,H_m,x_m).
\end{equation}
\end{proof}

\subsection{Technical Lemmas}
\label{app:theil-technical-lemmas}

\begin{lemma}[Marginal extrema of the summaries]
\label{lem:theil-summary-extrema}
For every integer $k\ge0$, the bounds satisfy
\begin{align}
\label{eq:app-theil-S-extrema}
    \underline S^{(k)}(X)
    &=\min_{\substack{X'\in[L,U]^n,\ d(X,X')\le k\\m\in\{1, \dots, n\}}}
          S_m(X'),
    &\overline S^{(k)}(X)
    &=\max_{\substack{X'\in[L,U]^n,\ d(X,X')\le k\\m\in\{1, \dots, n\}}}
          S_m(X'),\\
\label{eq:app-theil-H-extrema}
    \underline H^{(k)}(X)
    &=\min_{\substack{X'\in[L,U]^n,\ d(X,X')\le k\\m\in\{1, \dots, n\}}}
          H_m(X'),
    &\overline H^{(k)}(X)
    &=\max_{\substack{X'\in[L,U]^n,\ d(X,X')\le k\\m\in\{1, \dots, n\}}}
          H_m(X').
\end{align}
\end{lemma}

\begin{proof}
Let $q=q_k=\min\{k,n-1\}$. For any feasible $X'$ and any
$m\in\{1, \dots, n\}$, at most $q$ of the $n-1$ elements contributing to $S_m(X')$ can be changed. Hence, at least $n-q-1$ of these elements retain their original values. To minimize $S_m(X')$, we retain the $n-q-1$ smallest observations in $X$ and set the remaining $q$ values equal to $L$. Thus,
\begin{equation}
    S_m(X')
    \ge
    \sum_{i=1}^{n-q-1}x_{(i)}+qL
    =
    \underline S^{(k)}(X).
\end{equation}
This bound is attainable by choosing $m$ among the remaining
$q+1$ observations and replacing the other $q$ with $L$.
Similarly, retaining the $n-q-1$ largest observations and replacing the other $q$ values with $U$ gives
\begin{equation}
    S_m(X')
    \le
    \sum_{i=q+2}^{n}x_{(i)}+qU
    =
    \overline S^{(k)}(X),
\end{equation}
and this upper bound is also attainable. For $H_m(X')$, apply the same argument to $\phi(x)=x\log x$. Since
\begin{equation}
    \phi'(x)=1+\log x,
    \qquad
    \phi''(x)=\frac1x>0,
\end{equation}
the minimum of $\phi$ on $[L,U]$ is $\phi_{\min}$ and the maximum is $\phi_{\max}$ as defined in
\eqref{eq:theil-phi-extrema-main}. Therefore
\begin{equation}
    \underline H^{(k)}(X)
    =
    \sum_{i=1}^{n-q-1}v_{(i)}+q\phi_{\min}
    \le H_m(X')
    \le
    \sum_{i=q+2}^{n}v_{(i)}+q\phi_{\max}
    =
    \overline H^{(k)}(X).
\end{equation}
Both bounds are attainable by the same construction, using a
minimizer or maximizer of $\phi$ for the $q$ modified observations. Thus all four quantities are the exact marginal extrema.
\end{proof}

\begin{lemma}[Bounds on the one-dimensional minimizer]
\label{lem:theil-minimizer-bounds}
The minimum and maximum of $H/S$ over
$\mathcal C^{(k)}(X)$ are attained at the corners of the rectangle.
Moreover, for every $X'$ with $d(X,X')\le k$ and every $m\in\{1, \dots, n\}$,
\begin{equation}
    \underline t_T^{(k)}(X)
    \le
    t_T^\star(S_m(X'),H_m(X'))
    \le
    \overline t_T^{(k)}(X).
\end{equation}
\end{lemma}

\begin{proof}
Since $S\ge (n-1)L>0$, the ratio $H/S$ is increasing in $H$ for fixed $S$ and monotone in $S$ for fixed $H$. Hence, its minimum and maximum over the rectangle $\mathcal C^{(k)}(X)$ occur at one of the four corners. By Lemma~\ref{lem:theil-summary-extrema}, every feasible pair
$(S_m(X'),H_m(X'))$ belongs to $\mathcal C^{(k)}(X)$. Therefore,
\begin{equation}
    \underline\eta^{(k)}(X)
    \le
    \frac{H_m(X')}{S_m(X')}
    \le
    \overline\eta^{(k)}(X).
\end{equation}
Since the exponential function is increasing,
\begin{equation}
    \exp(\underline\eta^{(k)}(X))
    \le
    t_T^\star(S_m(X'),H_m(X'))
    \le
    \exp(\overline\eta^{(k)}(X)).
\end{equation}
Combining this with $t_T^\star(S_m(X'),H_m(X'))\in[L,U]$
gives the stated bounds $\underline t_T^{(k)}(X)$ and $\overline t_T^{(k)}(X)$.
\end{proof}

\subsection{Proof of Global Sensitivity}
\label{app:theil-global-sensitivity}

\renewcommand{\restatename}{Proposition~\ref{prop:theil-global-sensitivity}}
\begin{restate}[Global sensitivity]
When neighboring databases are in $[L,U]^n$,
\begin{equation}
    \operatorname{GS}_T
    =
    \frac{U}{(n-1)L+U}
    \log\left(\frac UL\right)
    -
    \log\left(
        \frac{(n-1)L+U}{nL}
    \right).
    \tag{\ref{eq:theil-global-sensitivity-LU}}
\end{equation}
The maximum is attained by the neighboring datasets $(L,\ldots,L)$ and $(U,L,\ldots,L)$. For $n=2$, the pair
$(U,U)$ and $(L,U)$ attains the same maximum.
\end{restate}

\begin{proof}
For any $X\in[L,U]^n$ and $m\in\{1, \dots, n\}$, write $t_m^\star=t_T^\star(S_m,H_m)$. Since $x_m$ can vary freely over $[L,U]$ without changing $(S_m,H_m)$, \eqref{eq:theil-R-main} gives
\begin{equation}
    \operatorname{GS}_T
    =
    \max_{X\in[L,U]^n}\max_{m\in\{1, \dots, n\}}
    R_T(S_m,H_m)
    =
    \max_{X,m}
    \max\left\{
        f_{T,L}(S_m,t_m^\star),
        f_{T,U}(S_m,t_m^\star)
    \right\}.
\end{equation}
We bound the two endpoint terms separately.

\paragraph{Upper endpoint.}
Let
\begin{equation}
    a=\frac{S_m}{n-1}\in[L,U].
\end{equation}
By convexity of $x\log x$,
\begin{equation}
    \frac{H_m}{n-1}\ge a\log a,
\end{equation}
and hence
\begin{equation}
    \frac{H_m}{S_m}\ge\log a,
    \qquad
    t_m^\star\ge a.
\end{equation}
Since $f_{T,U}(S,t)$ is nonincreasing in $t$,
\begin{equation}
    f_{T,U}(S_m,t_m^\star)
    \le
    f_{T,U}((n-1)a,a).
\end{equation}
Moreover,
\begin{equation}
\frac{d}{da}
\left[
    \frac{U}{(n-1)a+U}\log\left(\frac Ua\right)
    +
    \log\left(\frac{na}{(n-1)a+U}\right)
\right]
=
-\frac{(n-1)U\log(U/a)}
       {((n-1)a+U)^2}
\le0.
\end{equation}
Therefore, the largest value occurs at $a=L$, giving
\begin{equation}
\label{eq:app-theil-GS-upper}
    f_{T,U}(S_m,t_m^\star)
    \le
    \frac{U}{(n-1)L+U}\log\left(\frac UL\right)
    -
    \log\left(\frac{(n-1)L+U}{nL}\right).
\end{equation}

\paragraph{Lower endpoint.}
Since $H_m/S_m$ is a weighted average of $\{\log x_i:i\ne m\}$, we have $ t_m^\star \le \max_{i\ne m}x_i$. Because the remaining $n-2$ observations are at least $L$,
\begin{equation}
    S_m\ge t_m^\star+(n-2)L.
\end{equation}
For fixed $t\in[L,U]$, $f_{T,L}(S,t)$ is nonincreasing in
$S$ whenever $S\ge t$. To see this, let $v=t/L\ge1$. Then
\begin{equation}
    \frac{\partial}{\partial S}f_{T,L}(S,t)
    =
    \frac{
        L\{(\log v+1-v)S+L(v\log v-v+1)\}
    }
    {(S+L)^2(S+t)}.
\end{equation}
Since $\log v+1-v\le0$ and $S\ge t=Lv$, the numerator is at most
\begin{equation}
    L^2(2v\log v-v^2+1)\le0.
\end{equation}
Hence
\begin{equation}
    f_{T,L}(S_m,t_m^\star)
    \le
    f_{T,L}(t_m^\star+(n-2)L,t_m^\star).
\end{equation}
Now let $v=t_m^\star/L\in[1, U/L]$. Substitution gives
\begin{equation}
    f_{T,L}(t_m^\star+(n-2)L,t_m^\star)
    =
    \log\left(\frac{n-2+2v}{n-1+v}\right)
    -
    \frac{\log v}{n-1+v}.
\end{equation}
For $v\ge1$,
\begin{equation}
    \log\left(\frac{n-2+2v}{n-1+v}\right)
    -
    \frac{\log v}{n-1+v}
    \le
    \frac{v}{n-1+v}\log v
    -
    \log\left(\frac{n-1+v}{n}\right).
\end{equation}
The difference between the right- and left-hand sides
is zero at $v=1$, and its derivative is
\begin{equation}
    \frac{n-2}{(n-1+v)^2}
    \left[
        \log v
        -
        \frac{(v-1)(n-1+v)}
             {v(n-2+2v)}
    \right]
    \ge0,
\end{equation}
since
\begin{equation}
    \log v\ge\frac{v-1}{v},
    \qquad
    \frac{n-1+v}{n-2+2v}\le1.
\end{equation}
Finally,
\begin{equation}
    \frac{d}{dv}
    \left[
        \frac{v}{n-1+v}\log v
        -
        \log\left(\frac{n-1+v}{n}\right)
    \right]
    =
    \frac{(n-1)\log v}{(n-1+v)^2}
    \ge0.
\end{equation}
Since $v\le U/L$,
\begin{equation}
\label{eq:app-theil-GS-lower}
    f_{T,L}(S_m,t_m^\star)
    \le
    \frac{U}{(n-1)L+U}\log\left(\frac UL\right)
    -
    \log\left(\frac{(n-1)L+U}{nL}\right).
\end{equation}
To show that the bound is attained, consider $X_0=(L,\ldots,L)$ and $X_1=(U,L,\ldots,L)$. Since $T(X_0)=0$,
\begin{equation}
    T(X_1)
    =
    \frac{U}{(n-1)L+U}\log\left(\frac UL\right)
    -
    \log\left(\frac{(n-1)L+U}{nL}\right).
\end{equation}
Thus equality holds, proving
\eqref{eq:theil-global-sensitivity-LU}.
For $n=2$, the pair $(U,U)$ and $(L,U)$ attains the same
maximum as well.
\end{proof}

\subsection{Proof of the Neighborhood Upper Bound}
\label{app:theil-neighborhood-proof}

\renewcommand{\restatename}{Theorem~\ref{thm:theil-neighborhood-bound}}
\begin{restate}[A neighborhood upper bound for the Theil index]
For every $X\in[L,U]^n$ and every integer $k\ge0$,
$A_T^{(k)}(X)\le B_T^{(k)}(X)$.
\end{restate}

\begin{proof}
By Proposition~\ref{prop:theil-local-sensitivity},
\begin{equation}
\label{eq:app-theil-Ak-interchange}
    A_T^{(k)}(X)
    =
    \max_{\substack{X'\in[L,U]^n,\ d(X,X')\le k\\m\in\{1, \dots, n\}}}
    D_T(S_m(X'),H_m(X'),x_m').
\end{equation}
Fix any feasible $X'$ and $m$. Since
\begin{equation}
    D_T(S_m(X'),H_m(X'),x_m')
    \le
    R_T(S_m(X'),H_m(X')),
\end{equation}
and $h_T(\cdot;S_m(X'),H_m(X'))$ is minimized at $t_T^\star(S_m(X'),H_m(X'))$, we have
\begin{equation}
\begin{aligned}
    R_T(S_m(X'), H_m(X')) = \max \Bigl\{ & f_{T,L}\big(S_m(X'), t_T^\star(S_m(X'), H_m(X'))\big), \\
    & f_{T,U}\big(S_m(X'), t_T^\star(S_m(X'), H_m(X'))\big) \Bigr\}.
\end{aligned}
\end{equation}
By Lemma~\ref{lem:theil-minimizer-bounds},
\begin{equation}
    \underline t_T^{(k)}(X)
    \le
    t_T^\star(S_m(X'),H_m(X'))
    \le
    \overline t_T^{(k)}(X).
\end{equation}
Since $f_{T,L}(S,t)$ is nondecreasing in $t$, whereas
$f_{T,U}(S,t)$ is nonincreasing,
\begin{equation}
    f_{T,L}(S,t)
    \le
    f_{T,L}(S,\overline t_T^{(k)}(X)),
\end{equation}
and
\begin{equation}
    f_{T,U}(S,t)
    \le
    f_{T,U}(S,\underline t_T^{(k)}(X)).
\end{equation}
Moreover,
\begin{equation}
    S\in
    [\underline S^{(k)}(X),\overline S^{(k)}(X)].
\end{equation}
Therefore,
\begin{equation}
    D_T(S_m(X'),H_m(X'),x_m')
    \le
    \widetilde B_T^{(k)}(X).
\end{equation}
Taking the maximum over all feasible $X'$ and $m$ gives
\begin{equation}
    A_T^{(k)}(X)
    \le
    \widetilde B_T^{(k)}(X).
\end{equation}
Also, by definition of global sensitivity,
\begin{equation}
    A_T^{(k)}(X)
    \le
    \operatorname{GS}_T.
\end{equation}
Hence
\begin{equation}
    A_T^{(k)}(X)
    \le
    \min\left\{
        \operatorname{GS}_T,\,
        \widetilde B_T^{(k)}(X)
    \right\}
    =
    B_T^{(k)}(X),
\end{equation}
which proves the result.
\end{proof}

\subsection{Proof of the Smooth-Upper-Bound Property}
\label{app:theil-smooth-proof}

\renewcommand{\restatename}{Theorem~\ref{thm:theil-smooth-bound}}
\begin{restate}[A smooth upper bound for the Theil index]
For every $\beta>0$, $\overline{\mathcal S}_{T,\beta}$ is a $\beta$-smooth upper bound on the local sensitivity of the Theil index. In particular,
\begin{equation}
    \overline{\mathcal S}_{T,\beta}(X)
    \ge
    \mathcal S_{T,\beta}(X)
    \ge
    \operatorname{LS}_T(X),
    \tag{\ref{eq:theil-smooth-domination-main}}
\end{equation}
and, whenever $d(X,Y)=1$,
\begin{equation}
    \overline{\mathcal S}_{T,\beta}(X)
    \le
    e^\beta
    \overline{\mathcal S}_{T,\beta}(Y).
    \tag{\ref{eq:theil-smoothness-main}}
\end{equation}
\end{restate}

\begin{proof}
Let $X$ and $Y$ be neighboring datasets. If $d(X,X')\le k$, then the triangle inequality gives
\begin{equation}
    d(Y,X')\le k+1.
\end{equation}
Since the summary bounds in Lemma~\ref{lem:theil-summary-extrema} are exact marginal extrema over these neighborhoods and all excluded values,
\begin{align*}
    [\underline S^{(k)}(X),\overline S^{(k)}(X)]
    &\subseteq
    [\underline S^{(k+1)}(Y),\overline S^{(k+1)}(Y)],\\
    [\underline H^{(k)}(X),\overline H^{(k)}(X)]
    &\subseteq
    [\underline H^{(k+1)}(Y),\overline H^{(k+1)}(Y)].
\end{align*}
Hence,
\begin{equation}
\label{eq:app-theil-rectangle-nesting}
    \mathcal C^{(k)}(X)
    \subseteq
    \mathcal C^{(k+1)}(Y).
\end{equation}
Taking the minimum and maximum of $H/S$ over the nested
rectangles gives
\begin{equation}
    \underline\eta^{(k+1)}(Y)
    \le
    \underline\eta^{(k)}(X),
    \qquad
    \overline\eta^{(k)}(X)
    \le
    \overline\eta^{(k+1)}(Y),
\end{equation}
and therefore
\begin{equation}
\label{eq:app-theil-t-nesting}
    [\underline t_T^{(k)}(X),\overline t_T^{(k)}(X)]
    \subseteq
    [\underline t_T^{(k+1)}(Y),\overline t_T^{(k+1)}(Y)].
\end{equation}
Since $f_{T,L}(S,t)$ is nondecreasing in $t$, while the interval over which $S$ is maximized can only expand,
\begin{equation}
    B_{T,L}^{(k)}(X)
    \le
    B_{T,L}^{(k+1)}(Y).
\end{equation}
Similarly, because $f_{T,U}(S,t)$ is nonincreasing in $t$,
\begin{equation}
    B_{T,U}^{(k)}(X)
    \le
    B_{T,U}^{(k+1)}(Y).
\end{equation}
Consequently,
\begin{equation}
    \widetilde B_T^{(k)}(X)
    \le
    \widetilde B_T^{(k+1)}(Y);
\end{equation}
taking the minimum with the fixed constant
$\operatorname{GS}_T$ preserves the inequality. Hence,
\begin{equation}
\label{eq:app-theil-B-shift}
    B_T^{(k)}(X)
    \le
    B_T^{(k+1)}(Y),
    \qquad k\ge0.
\end{equation}
For every $k\ge n-1$, we have $q_k=n-1$, so
\begin{align*}
    \underline S^{(k)}(X)&=(n-1)L,
    &\overline S^{(k)}(X)&=(n-1)U,\\
    \underline H^{(k)}(X)&=(n-1)\phi_{\min},
    &\overline H^{(k)}(X)&=(n-1)\phi_{\max}.
\end{align*}
Thus, the corresponding bounds on $t_T^\star$ cover the full interval $[L,U]$. In particular,
the upper-endpoint bound includes
\begin{equation}
    f_{T,U}((n-1)L,L)
    =
    \operatorname{GS}_T.
\end{equation}
Therefore,
\begin{equation}
    \widetilde B_T^{(k)}(X)
    \ge
    \operatorname{GS}_T,
    \qquad k\ge n-1,
\end{equation}
and hence, by the definition of $B_T^{(k)}$,
\begin{equation}
\label{eq:app-theil-saturation}
    B_T^{(k)}(X)
    =
    \operatorname{GS}_T,
    \qquad k\ge n-1.
\end{equation}
Since $e^{-\beta k}$ decreases with $k$, no value of $k$ larger than
$n-1$ can improve the maximum. Thus,
\begin{equation}
    \overline{\mathcal S}_{T,\beta}(X)
    =
    \max_{k\ge0}
    e^{-\beta k}B_T^{(k)}(X)
    =
    \max_{0\le k\le n-1}
    e^{-\beta k}B_T^{(k)}(X).
\end{equation}
By Theorem~\ref{thm:theil-neighborhood-bound},
\begin{equation}
    A_T^{(k)}(X)
    \le
    B_T^{(k)}(X)
\end{equation}
for every $k$. Therefore, using the generic representation of
smooth sensitivity in \eqref{eq:generic-smooth-sensitivity},
\begin{equation}
    \mathcal S_{T,\beta}(X)
    \le
    \overline{\mathcal S}_{T,\beta}(X).
\end{equation}
Since the exact smooth sensitivity dominates local sensitivity,
\begin{equation}
    \overline{\mathcal S}_{T,\beta}(X)
    \ge
    \mathcal S_{T,\beta}(X)
    \ge
    \operatorname{LS}_T(X).
\end{equation}
Finally, for neighboring $X$ and $Y$,
\eqref{eq:app-theil-B-shift} gives
\begin{align}
    \overline{\mathcal S}_{T,\beta}(X)
    &=
    \max_{k\ge0}
    e^{-\beta k}B_T^{(k)}(X) \nonumber\\
    &\le
    \max_{k\ge0}
    e^{-\beta k}B_T^{(k+1)}(Y) \nonumber\\
    &=
    e^\beta
    \max_{j\ge1}
    e^{-\beta j}B_T^{(j)}(Y) \nonumber\\
    &\le
    e^\beta
    \overline{\mathcal S}_{T,\beta}(Y).
\end{align}
Thus $\overline{\mathcal S}_{T,\beta}$ is a
$\beta$-smooth upper bound on $\operatorname{LS}_T$.
\end{proof}

\section{Private Upper Bound Selection}
\label{sec:private-upper-bound}
The differentially private algorithms for the MLD and Theil index require an upper bound $U$ on the data values.  Setting $U$ to be very large can increase the smooth sensitivity (as well as the global and local sensitivity).  We therefore want to set $U$ close to the maximum value in $X$.  However, we cannot simply set $U$ to this maximum, as that would violate the DP guarantee.  Instead, we select $U$ using a differentially private algorithm, as we now describe.  In particular, we select $U$ with the unbounded quantile mechanism of
\citet{durfee2023unbounded}. This mechanism searches over a  geometric grid of candidate
thresholds using an algorithm called ``AboveThreshold'' \citep{dwork2014algorithmic}. In this appendix, we describe
the mechanism for selecting $U$ and its privacy guarantee, and we give a shortcut
for the unbounded search. Algorithm \ref{alg:private-upper-bound} summarizes the procedure at the end of this appendix.

Throughout, $\operatorname{Exp}(\lambda)$ denotes
the exponential distribution with rate $\lambda$, and
$\operatorname{Geom}(p)$ denotes the geometric distribution on
$\{1,2,\ldots\}$ with $\Pr(G=g)=(1-p)^{g-1}p$. The sample size $n$ is public.

\subsection{Mechanism and Privacy}
\label{subsec:upper-mechanism}

Fix a growth factor $\rho>1$. Define the candidate thresholds $U_j=L+\rho^j-1$ and the corresponding counts
$c_j(X)=\sum_{i=1}^n \mathbf{1}\{x_i<U_j\}$,
for $j=0,1,\ldots$.
Set $\varepsilon_\tau=\varepsilon_Q=\varepsilon_U/2$, and draw a noisy target
\begin{equation}
\label{eq:upper-target}
    \widetilde\tau=n+\frac{1}{\varepsilon_U}+E_\tau,
    \qquad
    E_\tau\sim\operatorname{Exp}(\varepsilon_\tau).
\end{equation}
Following the recommendation of \citet[Section~6.4]{durfee2023unbounded} for
heavy-tailed data, the target is placed above $n$ to encourage the search  to terminate close to the sample maximum. For $j=0,1,\ldots$, draw
$E_{Q,j}\sim\operatorname{Exp}(\varepsilon_Q)$ independently of each other and
of $E_\tau$, and let
\begin{equation}
\label{eq:upper-crossing}
    J=\min\{j\ge0: c_j(X)+E_{Q,j}\ge\widetilde\tau\},
    \qquad
    \widetilde U=U_J.
\end{equation}

\begin{proposition}[Privacy of upper bound selection]
\label{prop:upper-privacy}
The threshold $\widetilde U$ in \eqref{eq:upper-crossing} satisfies
$\varepsilon_U$-differential privacy under replacement adjacency.
\end{proposition}

\begin{proof}
Under replacement adjacency, the counts $\{c_j\}$ have sensitivity one and
are monotone, in the sense that for any neighboring $X,X'$ either
$c_j(X)\le c_j(X')$ for all $j$ or $c_j(X)\ge c_j(X')$ for all $j$
\citep[Lemma~4.1]{durfee2023unbounded}. The procedure
\eqref{eq:upper-target}--\eqref{eq:upper-crossing} is AboveThreshold with
exponential noise and a data-independent threshold $n+1/\varepsilon_U$, which
is $(\varepsilon_\tau+\varepsilon_Q)$-differentially private for monotone
queries \citep[Corollary~3.1]{durfee2023unbounded}. Since
$\widetilde U=U_J$ is a deterministic function of $J$, the result follows by
post-processing.
\end{proof}

\subsection{Exact Acceleration of the Saturated Tail}
\label{subsec:upper-tail-acceleration}

Evaluating \eqref{eq:upper-crossing} directly may require many comparisons beyond the largest observation. Let
\begin{equation}
\label{eq:upper-jsat}
    j_{\mathrm{sat}}=\min\{j\ge0:U_j>\max_i x_i\}.
\end{equation}
For every $j\ge j_{\mathrm{sat}}$, $c_j(X)=n$, so conditional on $\widetilde\tau$ each such comparison succeeds with probability
\begin{equation}
\label{eq:upper-tail-prob}
    p=\Pr(E_{Q,j}\ge\widetilde\tau-n\mid\widetilde\tau)
    =\exp\{-\varepsilon_Q(\widetilde\tau-n)\}.
\end{equation}
Because $\widetilde\tau-n\ge1/\varepsilon_U>0$, we have $0<p<1$.

\begin{lemma}[Exact tail acceleration]
\label{lem:upper-tail-acceleration}
Suppose no crossing occurs at $j<j_{\mathrm{sat}}$. Replacing the comparisons
at $j\ge j_{\mathrm{sat}}$ by a draw $G\mid\widetilde\tau\sim
\operatorname{Geom}(p)$ and setting $J=j_{\mathrm{sat}}+G-1$ leaves the
distribution of $J$, and hence of $\widetilde U$, unchanged. In particular,
$J<\infty$ almost surely.
\end{lemma}

\begin{proof}
Conditional on $\widetilde\tau$ and on no crossing before
$j_{\mathrm{sat}}$, the indicators
$\mathbf 1\{n+E_{Q,j}\ge\widetilde\tau\}$, $j\ge j_{\mathrm{sat}}$, are
independent Bernoulli$(p)$ variables, because the $E_{Q,j}$ are independent
of $\widetilde\tau$ and of the earlier draws. The number of comparisons up to
and including the first success is therefore $\operatorname{Geom}(p)$, which
is finite almost surely since $p>0$. Integrating over $\widetilde\tau$ and
over the comparisons at $j<j_{\mathrm{sat}}$ gives the result.
\end{proof}

Although $j_{\mathrm{sat}}$ depends on $\max_i x_i$, it affects only how $J$ is computed, not its distribution. The output has exactly the distribution of the
mechanism in \eqref{eq:upper-crossing}, so
Proposition~\ref{prop:upper-privacy} applies unchanged. 

\begin{algorithm}[t]
\caption{Private upper bound selection}
\label{alg:private-upper-bound}
\begin{algorithmic}[1]
\Require Data $X=(x_1,\ldots,x_n)$; public lower bound $L$; growth factor $\rho>1$; privacy budget $\varepsilon_U$
\Ensure Private clipping threshold $\widetilde U$, which is $\varepsilon_U$-differentially private
\State $\varepsilon_\tau\gets\varepsilon_U/2$,\quad $\varepsilon_Q\gets\varepsilon_U/2$;\quad $U_j\gets L+\rho^j-1$
\State Draw $E_\tau\sim\operatorname{Exp}(\varepsilon_\tau)$;\quad $\widetilde\tau\gets n+1/\varepsilon_U+E_\tau$
\State $j_{\mathrm{sat}}\gets\min\{j\ge0:U_j>\max_i x_i\}$
\For{$j=0,\ldots,j_{\mathrm{sat}}-1$}
    \State $c_j\gets\sum_{i=1}^n\mathbf 1\{x_i<U_j\}$
    \State Draw $E_{Q,j}\sim\operatorname{Exp}(\varepsilon_Q)$
    \If{$c_j+E_{Q,j}\ge\widetilde\tau$}
        \State \Return $\widetilde U\gets U_j$
    \EndIf
\EndFor
\Statex \textit{Reached only if no crossing occurs before $j_{\mathrm{sat}}$:}
\State $p\gets\exp\{-\varepsilon_Q(\widetilde\tau-n)\}$ 
\State Draw $G\sim\operatorname{Geom}(p)$
\State \Return $\widetilde U\gets U_{j_{\mathrm{sat}}+G-1}$
\end{algorithmic}
\end{algorithm}

\end{appendices}

\end{document}